\documentclass[twocolumn]{aastex631}

\usepackage{comment}
\usepackage{soul}

\newcommand\E[1]{\cdot10^{#1}}
\newcommand\rs[1]{_\mathrm{#1}}
\def\gsim{\;\lower4pt\hbox{${\buildrel\displaystyle>\over\sim}$}\,}
\def\lsim{\;\lower4pt\hbox{${\buildrel\displaystyle<\over\sim}$}\,}
\begin{document}

\title{
Time evolution of the synchrotron X-ray emission in Kepler's SNR: the effects of turbulence and shock velocity
}

\author[0000-0002-6045-136X]{Vincenzo Sapienza}
\affiliation{Dipartimento di Fisica e Chimica E. Segr\`e, Universit\`a degli Studi di Palermo, Piazza del Parlamento 1, 90134, Palermo, Italy}
\affiliation{INAF-Osservatorio Astronomico di Palermo, Piazza del Parlamento 1, 90134, Palermo, Italy}

\author[0000-0003-0876-8391]{Marco Miceli}
\affiliation{Dipartimento di Fisica e Chimica E. Segr\`e, Universit\`a degli Studi di Palermo, Piazza del Parlamento 1, 90134, Palermo, Italy}
\affiliation{INAF-Osservatorio Astronomico di Palermo, Piazza del Parlamento 1, 90134, Palermo, Italy}

\author[0000-0003-3487-0349]{Oleh Petruk}
\affiliation{INAF-Osservatorio Astronomico di Palermo, Piazza del Parlamento 1, 90134, Palermo, Italy}
\affiliation{Institute for Applied Problems in Mechanics and Mathematics, Naukova Street 3-b, 79060 Lviv, Ukraine}
\affiliation{Astronomical observatory, Ivan Franko National University of Lviv, Kyryla i Methodia St. 8, UA-79005 Lviv, Ukraine}

\author[0000-0003-0890-4920]{Aya Bamba}
\affiliation{Department of Physics, Graduate School of Science, The University of Tokyo, 7-3-1 Hongo, Bunkyo-ku, Tokyo 113-0033, Japan}
\affiliation{Research Center for the Early Universe, School of Science, The University of Tokyo, 7-3-1 Hongo, Bunkyo-ku, Tokyo 113-0033, Japan}
\affiliation{Trans-Scale Quantum Science Institute, The University of Tokyo\\ 7-3-1 Hongo, Bunkyo-ku, Tokyo 113-0033, Japan}

\author[0000-0002-1104-7205]{Satoru Katsuda}
\affiliation{Graduate School of Science and Engineering, Saitama University, 255 Simo-Ohkubo, Sakura-ku, Saitama city, Saitama, 338-8570, Japan}

\author[0000-0003-2836-540X]{Salvatore Orlando}
\affiliation{INAF-Osservatorio Astronomico di Palermo, Piazza del Parlamento 1, 90134, Palermo, Italy}

\author[0000-0002-2321-5616]{Fabrizio Bocchino}
\affiliation{INAF-Osservatorio Astronomico di Palermo, Piazza del Parlamento 1, 90134, Palermo, Italy}

\author{Tracey DeLaney}
\affiliation{Physics and Engineering Department, West Virginia Wesleyan College, Buckhannon, WV 26201, USA}




\begin{abstract}

The maximum energy of electrons in supernova remnant (SNR) shocks is typically limited by radiative losses, where the synchrotron cooling time equals the acceleration time. 
The low speed of shocks in a dense medium increases the acceleration time, leading to lower maximum electron energies and fainter X-ray emissions. 
However, in Kepler's SNR, an enhanced electron acceleration, which proceeds close to the Bohm limit, occurs in the north of its shell, where the shock is slowed by a dense circumstellar medium (CSM). 
To investigate whether this scenario still holds at smaller scales, we analyzed the temporal evolution of the X-ray synchrotron flux in filamentary structures, using the two deepest \textit{Chandra}/ACIS X-ray observations, performed in 2006 and 2014. 
We examined spectra from different filaments,
we measured their proper motion and calculated the acceleration to synchrotron time-scale ratios. 
The interaction with the turbulent and dense northern CSM induces competing effects on electron acceleration: on one hand, turbulence reduces the electron mean free path enhancing the acceleration efficiency, on the other hand, lower shock velocities increase the acceleration time-scale. 
In most filaments, these effects compensate each other, but in one region the acceleration time-scale exceeds the synchrotron time-scale, resulting in a significant decrease in nonthermal X-ray emission from 2006 to 2014, indicating fading synchrotron emission. 
Our findings provide a coherent understanding of the different regimes of electron acceleration observed in Kepler's SNR through various diagnostics.

\end{abstract}



\section{Introduction} \label{sect:intro}

Supernova Remnants (SNRs) are prominent accelerators of particles, thus they are widely considered the primary origin of galactic cosmic rays.
The first observational evidence of very high energy ($E > 10^{12}$ eV) electrons accelerated at SNR shocks was discovered by \citet{1995Natur.378..255K}, who detected nonthermal X-ray emission stemming from SN 1006 (where also efficient hadronic acceleration has been reported, e. g., \citealt{2022NatCo..13.5098G}).
Indeed, outer shell of young SNRs typically emit synchrotron radiation in the X-ray band, which can be used as a diagnostic tool to deepen our understanding of the acceleration process.
The study of X-ray synchrotron emission can provide information about the shape of the electron energy distribution and the mechanisms that limit the maximum energy that electrons can reach (e. g., \citealt{2013A&A...556A..80M}).

Kepler's SNR, the aftermath of the explosion of the historical SN1604, is an interesting object to study the acceleration process, in order to study how the environment affect the acceleration mechanism.
The remnant, stemming a type Ia SN \citep{1999PASJ...51..239K}, is indeed interacting with a dense nitrogen-rich cloud circumstellar medium (CSM) in the north (\citealt{1991ApJ...366..484B,2007ApJ...662..998B,2007ApJ...668L.135R,2008ApJ...689..225K,2015ApJ...808...49K,2021ApJ...915...42K}).
Recent estimates based on proper motion measurements derived a distance $d = 5.1_{-0.7}^{+0.8}$ kpc \citep{2016ApJ...817...36S}.
We will adopt $d=5$ kpc throughout this paper.

Prominent particle acceleration in Kepler's SNR is testified by its energetic non-thermal emission.
The presence of non-thermal X-ray emission in Kepler's SNR was first discovered in its south-eastern region by \cite{2004A&A...414..545C}.
Recently, \citet{2021PASJ...73..302N} reported the first strong detection of hard X-ray emission within the 15-30 keV band from Kepler's SNR, by analyzing a \emph{Suzaku} HXD observation.

In a recent study, \citep{2021ApJ...907..117T} analyzed the cutoff photon energy ($\varepsilon_0$) of synchrotron radiation across different regions of several SNRs, including Kepler's SNR. 
To this end, the authors used a model of synchrotron emission, originally proposed by \cite{2007A&A...465..695Z}, where the electron maximum energy is limited by radiation losses.
In this scenario, $\varepsilon_0$  is related to the shock speed, $v_{sh}$, as 
\begin{equation} 
\label{eq:eovsvsh}
    \varepsilon_0=\frac{1.6}{\eta}\bigg( \frac{v_{sh}}{4000\text{ km s}^{-1}}\bigg)^2 \text{keV},
\end{equation}
where $\eta$, or Bohm factor, is the ratio between the diffusion coefficient and $c\lambda/3$ (where $\lambda$ is the Larmor radius, the minimum value $\eta=1$ corresponding to the Bohm limit) and is strongly related to the turbulence of the magnetic field, which scatter the charged particles in the acceleration process.
The spatially resolved analysis of Kepler's SNR by \citet{2021ApJ...907..117T}, lacked of the hard part of the spectrum, so no clear correlation between the shock velocity and the synchrotron cutoff energy was found, except for the synchrotron dominated regions.

In \cite{2022ApJ...935..152S}, by making use of \textit{NuSTAR} and \textit{XMM-Newton} data, we performed a spatially resolved spectral analysis of Kepler's SNR, including the hard part of the X-ray spectrum where the emission is dominated by synchrotron radiation.
The spectra were analyzed by adopting the loss-limited model.
We identified two different regimes of particle acceleration, characterized by different Bohm factors.
In the north, where the shock interacts with a dense CSM, we found a more efficient acceleration (i.e., lower Bohm factor) than in the south, where the shock velocity is higher and there are no signs of shock interaction with dense CSM.
This can be explained by considering that the interaction of the shock front with the dense CSM at north is associated with a turbulent magnetic field, which boosts the particle acceleration process.
On the other hand, the low shock speeds measured in the north \citep{2022ApJ...926...84C} lead to a high acceleration time scale ($\sim$300 yrs), which can result in a decrease of the maximum electron energy by radiation losses.
To unravel this intricate scenario on smaller scales we studied the evolution of the synchrotron flux in several filaments of Kepler's SNR, making use of the two deepest sets of \textit{Chandra}/ACIS observations performed in 2006 and 2014.

The paper is organized as follows: in Section \ref{sect:data} we present the datasets and the data reduction process, whereas in Section \ref{sect:results} we show the results obtained with the spectral analysis. Discussions and Conclusion are drawn respectively in Section \ref{sect:disc} and Section \ref{sec:con}.

\section{Observation and Data Reduction}\label{sect:data}
For our analysis we made use of different \textit{Chandra}/ACIS observations for the two epochs 2006 and 2014, as summarized in Table \ref{tab:obs}.
\begin{table}
    \centering
    \caption{\textit{Chandra}/ACIS observations table.}
    \resizebox{\columnwidth}{!}{
    \begin{tabular}{ccccc}
    \hline\hline
    Obs ID & Exp. Time (ks) & R. A.                  & Dec.                      & Start Date \\
    \hline
    6714   & 157.8          & 17$^h$ 30$^m$ 42.0$^s$ & -21\textdegree 29' 00.0'' & 27/04/2006 \\
    6715   & 159.1          & 17$^h$ 30$^m$ 41.2$^s$ & -21\textdegree 29' 31.4'' & 03/08/2006 \\
    6716   & 158.0          & 17$^h$ 30$^m$ 42.0$^s$ & -21\textdegree 29' 00.0'' & 05/05/2006 \\
    6717   & 106.8          & 17$^h$ 30$^m$ 41.2$^s$ & -21\textdegree 29' 31.4'' & 13/07/2006 \\
    6718   & 107.8          & 17$^h$ 30$^m$ 41.2$^s$ & -21\textdegree 29' 31.4'' & 21/07/2006 \\
    7366   & 51.5           & 17$^h$ 30$^m$ 41.2$^s$ & -21\textdegree 29' 31.4'' & 16/07/2006 \\
    16004  & 102.7          & 17$^h$ 30$^m$ 41.2$^s$ & -21\textdegree 29' 31.4'' & 13/05/2014 \\
    16614  & 36.4           & 17$^h$ 30$^m$ 41.2$^s$ & -21\textdegree 29' 31.4'' & 16/05/2014 \\
    \hline
    \end{tabular}
    }
    \label{tab:obs}
\end{table}
The data were reprocessed with the CIAO v4.13 software \citep{2006SPIE.6270E..1VF} using CALDB 4.9.4.
We reprocessed the data by using the \texttt{chandra\_repro} task. 
We mosaicked flux images for each year by using the \texttt{merge\_obs} task.
In order to measure proper motion, we followed the same astrometric alignment procedure described by \cite{2022ApJ...926...84C}, selecting the same point sources (also highlighted in yellow in Figure \ref{fig:reg}).
For the proper motion measurement only, we use the deep 2006 observation (Obs. ID: 6715) as the relative reference to which we aligned the 2014 deepest observation (Obs. ID: 16004), in order to minimize systematic errors in the reprojections.
To extract spectra, we used \texttt{specextract} CIAO command.
We then combined the spectra from the same epochs by using the \texttt{combine\_spectra} CIAO command.
The spectra were binned using the optimal binning algorithm \citep{2016A&A...587A.151K}.
The spectral analysis was performed with XSPEC v. 12.11.1 \cite{1996ASPC..101...17A}. 
We adopted the Cash statistic (C-stat) for the fitting procedure.

\section{Results} \label{sect:results}
\subsection{Spectra}
\begin{figure}
    \centering
    \includegraphics[width=\columnwidth]{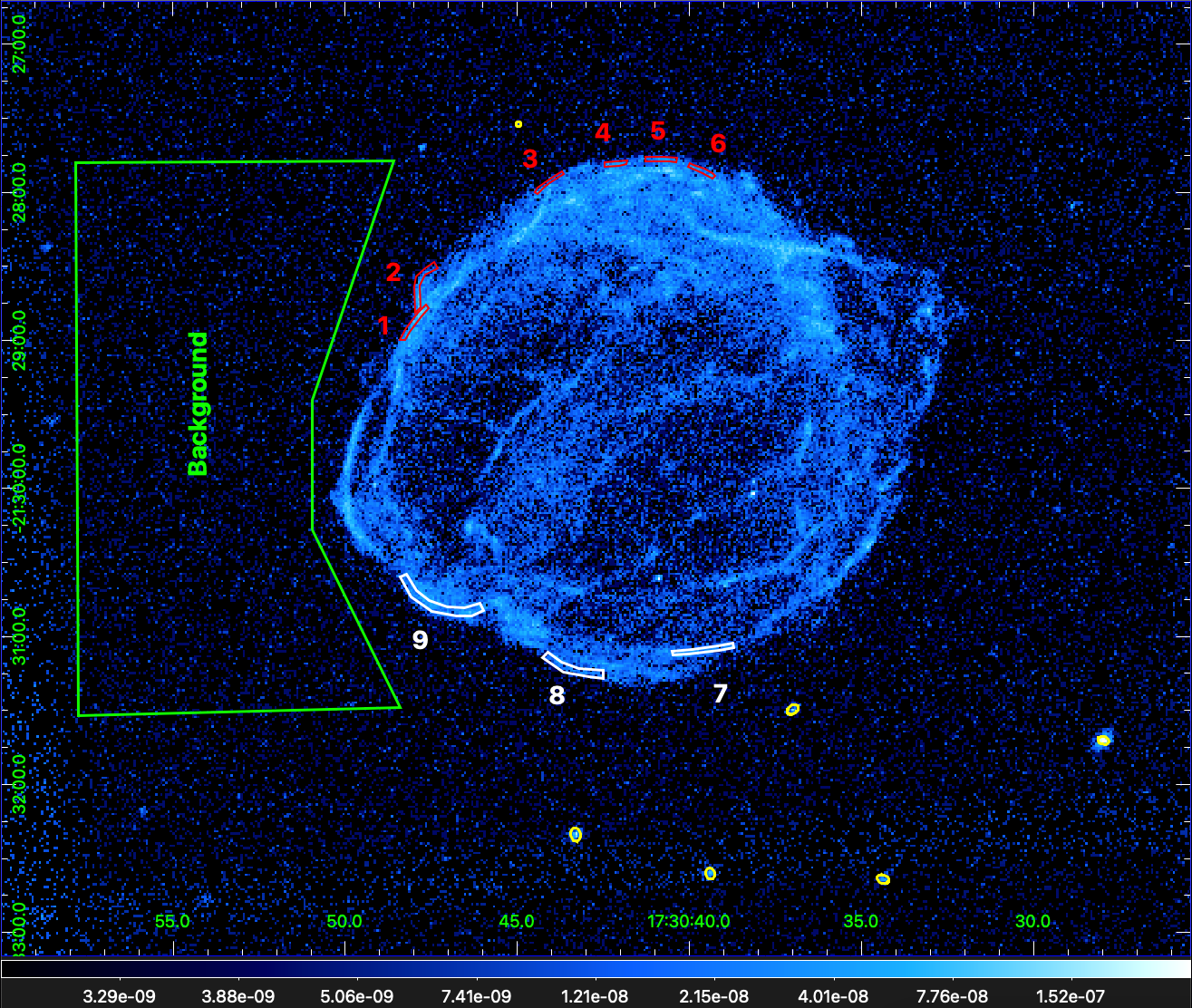}
    \caption{\textit{Chandra}/ACIS flux map of Kepler's SNR in $4.1-6$ keV band. 
    The colorbar is in logarithm scale in unit of photons cm$^{-2}$ s$^{-1}$.
    Source regions are marked with red polygons for the north and white polygons for the south.
    Background region is marked with the green polygon and yellow ellipses are the \cite{2022ApJ...926...84C} point sources used for the astrometric alignment.}
    \label{fig:reg}
\end{figure}
For our spatially resolved spectral analysis we consider only regions characterized by: i) a bright synchrotron emission, ii) a sharp edge in the radial distribution of surface brightness (see \citealt{2005ApJ...621..793B}), iii) low contamination from thermal emission and iv) at least 1200 photon counts in the spectrum of each epoch (photon counts for all regions are reported in Table \ref{tab:ecut} for the two epochs).
We identify 6 filamentary regions in the northern part of Kepler's SNR that are clearly visible in the $4.1-6$ keV energy band, were the synchrotron radiation dominates the emission, and are located at the rim of the shell. These regions are shown in red in Figure \ref{fig:reg} and labeled as $1-6$. 
As a comparison, we also consider 3 regions in the south (shown in white and labeled as $7-9$ in Figure \ref{fig:reg}).

The source spectra were fitted by adopting the loss-limited model proposed by \cite{2007A&A...465..695Z}, to describe the synchrotron emission, and a component from optically thin plasma in non-equilibrium of ionization (\texttt{vnei}) to account for the residual thermal emission (plus 2 Gaussian components to take in account the missing Fe L line in the model, see \citealt{2015ApJ...808...49K} and \citealt{2022ApJ...935..152S}). 
We modeled the background by a phenomenological fitting of the spectrum extracted from the green region in Figure \ref{fig:reg}.
The background model (properly scaled to account for the different areas of the extraction regions) was then added to the source model.
From the best-fit model of each region, we computed the flux in the $4.1-6$ keV band by using the \texttt{cflux} convolution model\footnote{In the \texttt{cflux} model, the normalization of one of the additive models must be fixed to a non-zero value (\url{https://heasarc.gsfc.nasa.gov/xanadu/xspec/manual/node289.html}), so we fixed the normalization of the thermal component.}.
We note that in the $4.1-6$ keV band the flux is dominated by the nonthermal component, whose contribution is always $>90\%$ of the total in each region (see also Figure \ref{fig:2006spectra} and \ref{fig:2014spectra} in Appendix).
The best-fit parameters for 2006 and 2014 are shown in Table \ref{tab:ecut}, while the spectra with the best-fit model and residuals are shown in Appendix in Figure \ref{fig:2006spectra} and Figure \ref{fig:2014spectra}.
\begin{table*}
    \centering
    \caption{Best-fit parameters for the regions shown in Figure \ref{fig:reg}. 
    Errors for $\varepsilon_0$ and flux are at the 68\% confidence level. Errors for shock velocity are at 90\% confidence level.} 
\begin{tabular}{cccccccc}

\hline\hline
Region \#&\multicolumn{2}{c}{$\varepsilon_0$ (keV)}&\multicolumn{2}{c}{Flux $4.1-6$ keV (Log$_{10}$ erg cm$^{-2}$ s$^{-1}$)}&V$_{sh}$ (km s$^{-1}$)&\multicolumn{2}{c}{Counts \# (0.5-8.0 keV)} \\
 & 2006 & 2014 & 2006 & 2014 & & 2006 & 2014\\
\hline 
1&$0.50_{-0.04}^{+0.04}$   &$0.9_{-0.2}^{+0.3}$   &$-13.742\pm0.014$&$-13.68\pm0.03$&$3570\pm100$&13023& 2072\\
2&$0.71_{-0.06}^{+0.07}$   &$0.80_{-0.16}^{+0.23}$&$-13.808\pm0.014$&$-13.81\pm0.03$&$4690\pm120$&8151&1364 \\
3&$0.40_{-0.05}^{+0.07}$   &$0.43_{-0.14}^{+0.28}$&$-14.36\pm0.03$  &$-14.33\pm0.07$&$3690\pm50$&13350& 1678 \\
4&$0.34_{-0.03}^{+0.04}$   &$0.38_{-0.09}^{+0.14}$&$-14.17\pm0.02$  &$-14.21\pm0.05$&$1870\pm70$&8778& 1238 \\
5&$0.35_{-0.04}^{+0.04}$   &$0.19_{-0.03}^{+0.05}$&$-14.19\pm0.02$  &$-14.42\pm0.06$&$1520\pm100$&9824 & 1494\\
6&$0.193_{-0.015}^{+0.017}$&$0.25_{-0.04}^{+0.06}$&$-14.22\pm0.02$  &$-14.20\pm0.06$&$1590\pm60$ &32217&4861\\
7&$0.32_{-0.02}^{+0.03}$   &$0.51_{-0.10}^{+0.16}$&$-13.963\pm0.018$&$-13.89\pm0.04$&$4160\pm70$&14717& 2246\\
8&$1.07_{-0.10}^{+0.12}$   &$0.89_{-0.16}^{+0.24}$&$-13.499\pm0.011$&$-13.51\pm0.03$&$7690\pm70$ &8133&1432 \\
9&$0.43_{-0.03}^{+0.04}$   &$0.36_{-0.05}^{+0.07}$&$-13.988\pm0.018$&$-13.96\pm0.04$&$6000\pm100$ &15631&2528\\
 \hline
\end{tabular}

\label{tab:ecut}
\end{table*}

\subsection{Shock velocity measurement}
To estimate of the shock velocity, we measured the proper motion from 2006 to 2014 in all the nine regions.
We mirrored the methodology adopted by \cite{2008ApJ...678L..35K}, where a deeper description of the procedure can be found. 
We extract the one dimensional radial count profiles of each filament from both 2006 and 2014 epochs.
The profiles were extracted using \textit{Chandra}/ACIS events file with 0.492" pixel.
Each profile was then remapped into a 40 times denser grid using a quadratic interpolation, in a similar fashion to \cite{2016ApJ...823L..32W}.
The square root of the counts was taken as statistical uncertainty.
We then shifted the 2014 profile relative to the 2006 profile, minimizing the value of $\chi^2$. 
\begin{figure*}[t!]
    \centering
    \includegraphics[width=0.49\textwidth]{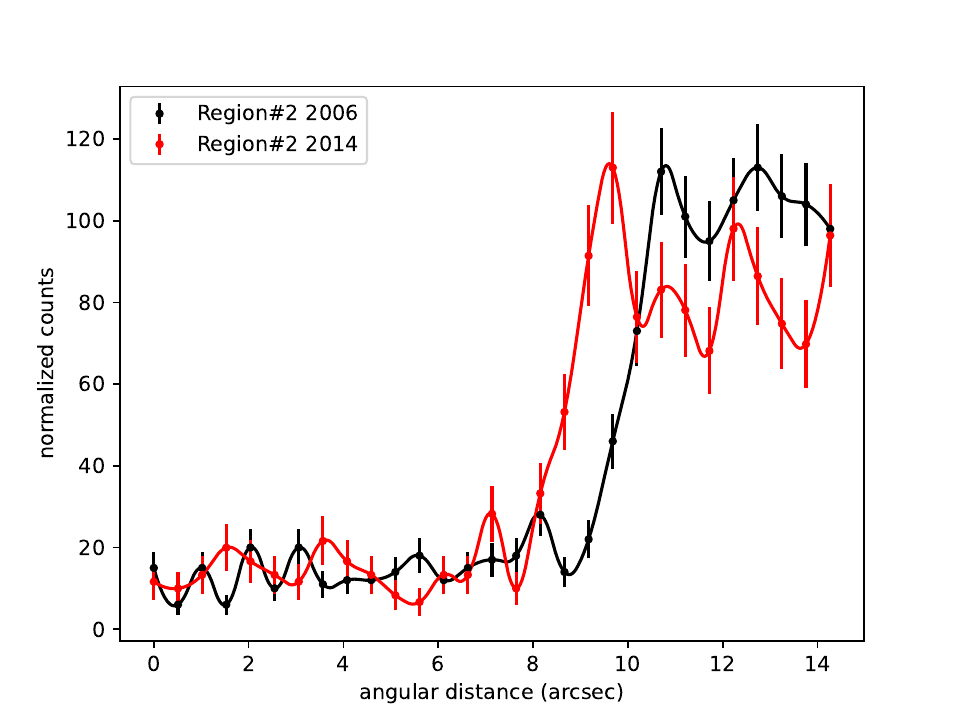}
    \includegraphics[width=0.49\textwidth]{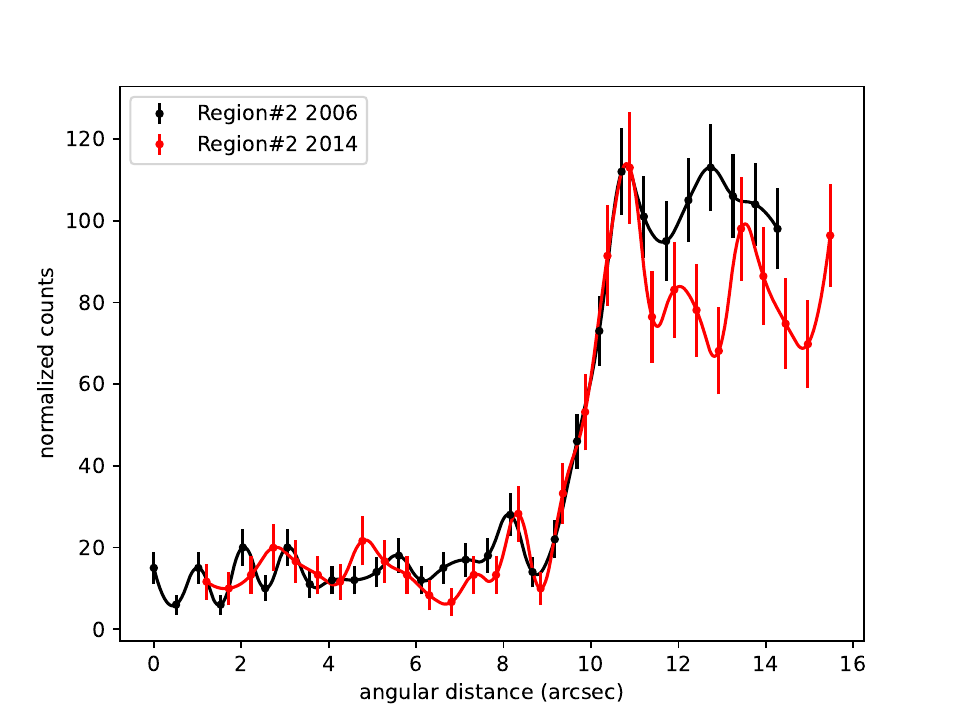}
    \caption{\emph{Left panel:} Example of the one dimensional radial count profile for region 2 in 2006 (in black) and in 2014 (in red). \emph{Right panel:} same as the left panel, but with the 2014 profile shifted according to the procedure described in \cite{2016ApJ...823L..32W}.}
    \label{fig:pmeg}
\end{figure*}
Figure \ref{fig:pmeg} shows the example of the radial count profiles (for region 2) before and after the shifting procedure.
In this paper, we only report the statistical errors, which are the 90\% confidence limits resulting from a $\chi^2$ increase $\Delta\chi^2=2.706$.
Once obtained the best-fit value for the angular shift ($\theta$), one can derive the shock velocities for each region (assuming a distance $d=5$ kpc, as stated in the Introduction). 
The shock velocities with their uncertainties are also reported in Table \ref{tab:ecut}.
The shock velocities displayed in Table \ref{tab:ecut} exhibit a large dispersion, despite the almost circular  appearance of the remnant.
Nonetheless, our measurements are in good agreement with previous studies (see \citealt{2008ApJ...689..225K} and \citealt{2022ApJ...926...84C}).
Similar velocity dispersions have been observed in other round-shaped SNRs such as SN 1006 \citep{2013ApJ...763...85K,2014ApJ...781...65W}, RCW 86 \citep{2016ApJ...820L...3Y,2022ApJ...938...59S} and Tycho's SNR \citep{2016ApJ...823L..32W}.

\section{Discussion} \label{sect:disc}
\subsection{Cut-off photon energy $\varepsilon_0$ vs. the shock speed $V_{sh}$}
\label{sect:cut-vsh}
\begin{figure*}
    \centering
    \includegraphics[width=0.49\textwidth]{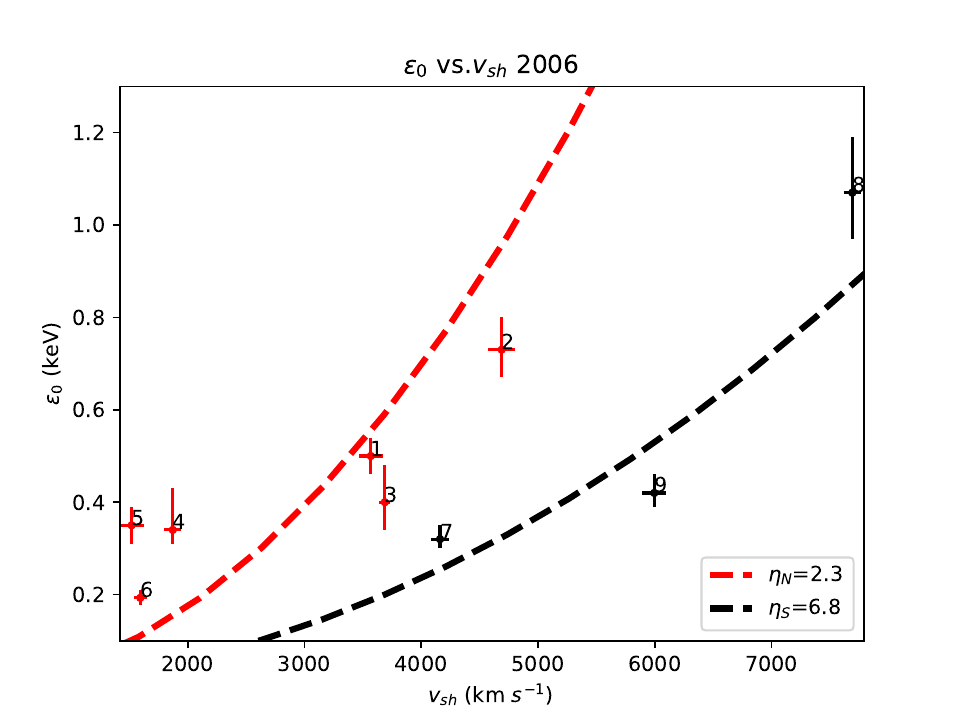}
    \includegraphics[width=0.49\textwidth]{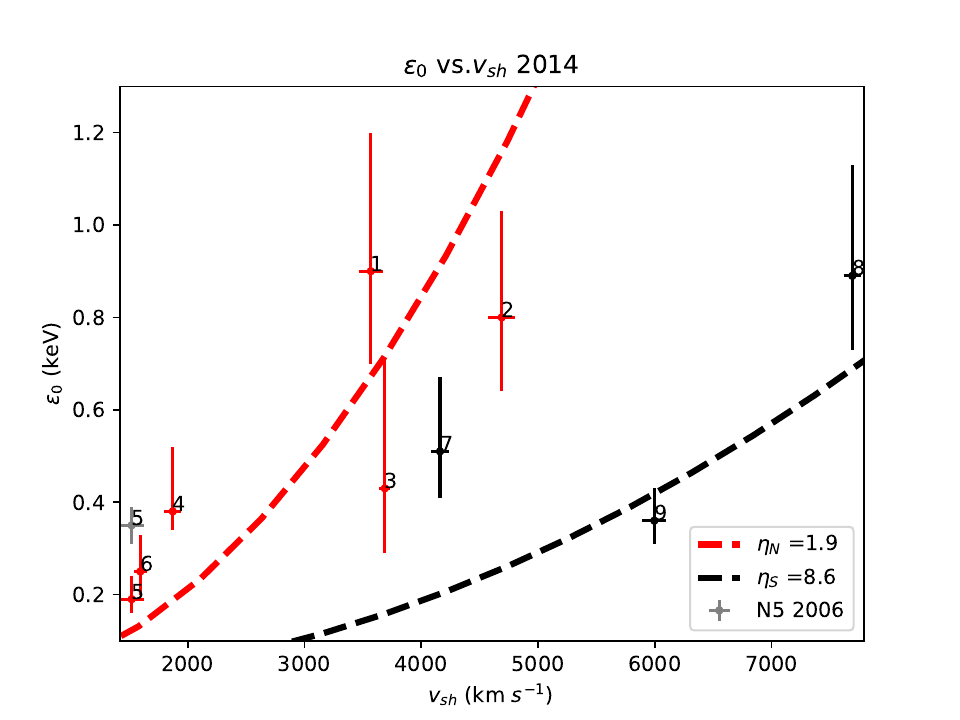}
    \caption{
    Synchrotron cutoff energy vs. current shock velocity for the year 2006 (\textit{left panel}) and for the year 2014 (\textit{right panel}).
    Red crosses mark northern regions (1-6) and the red dashed curve is the corresponding best-fit curve obtained from Equation \ref{eq:eovsvsh}. Black crosses mark southern regions (7-9) and the black dashed curve is the corresponding best-fit curve obtained from Equation \ref{eq:eovsvsh}.
    The gray cross in the right is the region 5 in 2006.
    }
    \label{fig:ecut}
\end{figure*}
By employing the same methodology outlined in our previous work, deeply described in \citet{2022ApJ...935..152S}, we present in Figure \ref{fig:ecut} the values of $\varepsilon_0$ for the two epochs analyzed: 2006 (left panel) and 2014 (right panel; listed in Table \ref{tab:ecut}) as a function of their corresponding shock velocity ($v_{sh}$).
We assumed the same shock velocity for the two epochs.
We use different colors to distinguish between data points derived from southern regions (in black) and northern regions (in red).
Notably, among all the regions examined, it is only region 5 that shows a significant decline in $\varepsilon_0$ between 2006 and 2014.
Figure \ref{fig:ecut} clearly illustrates the separation of the data points into two distinct clusters (in both epochs), representing the southern and northern regions. 
This result clearly confirms the existence of two distinct regimes of electron acceleration within the same SNR, already identified by \cite{2022ApJ...935..152S}.
By fitting each of these two clusters using Equation \ref{eq:eovsvsh}, we can derive the corresponding best-fit values of the Bohm  factor. 
For the southern regions, the retrieved Bohm factors are $\eta_S=6.8\pm 1.1$ in 2006 and $\overline{\eta}_S=8.6\pm 2.0$ in 2014. As for the northern regions, the corresponding Bohm factors are $\eta_N=2.3\pm 0.4$ in 2006 and $\overline{\eta}_N=1-3.7$ in 2014.
The best-fit values of the Bohm factors align well with those obtained in our previous research \citep{2022ApJ...935..152S}, further strengthening the evidence that the electron acceleration proceeds much closer to the Bohm limit in the north than in the south. 
In particular, $\eta_S/\eta_N=3.0\pm0.7$ and $\overline{\eta}_S/\overline{\eta}_N=3.6\pm2.0$ in 2006 and 2014, respectively.

\subsection{Synchrotron Losses vs. Acceleration Time Scale}
\label{sect:times}
\begin{figure}[t]
    \centering
    \includegraphics[trim={14pt 0 40pt 0},clip, width=\columnwidth]{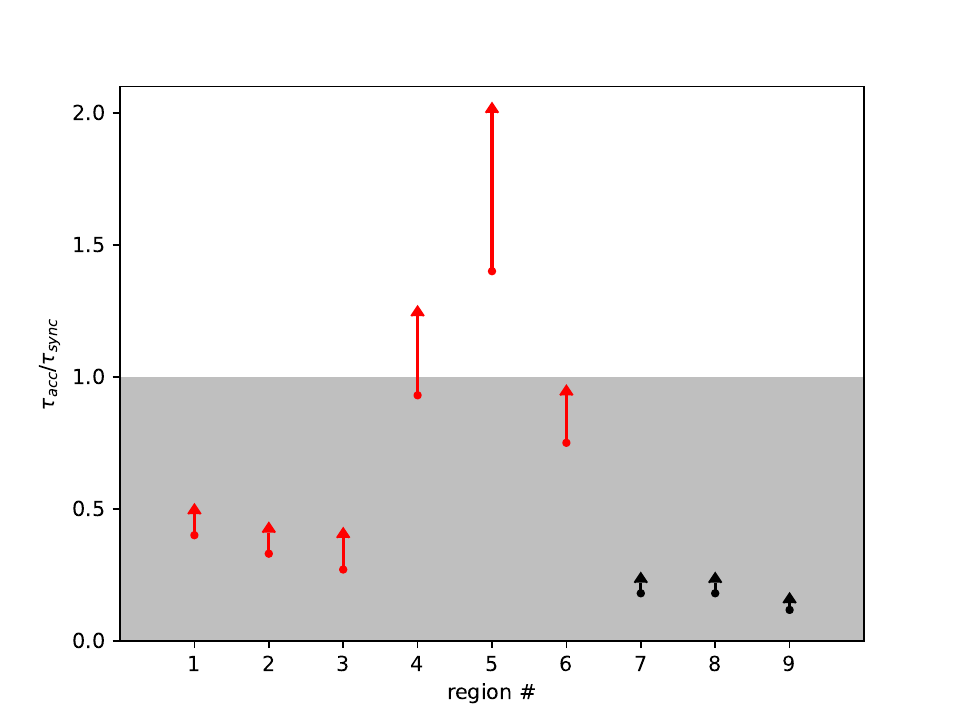}
    \caption{Lower limit of the ratio between the acceleration time scale and the synchrotron losses time scale, for the selected region.
    The arrows indicate the error associated to the lower limit value.}
    \label{fig:ratio}
\end{figure}

Comparing the acceleration time to the synchrotron losses time scale is crucial for determining if the acceleration mechanisms can effectively counteract particle losses, thereby indicating whether the region under investigation follows a loss-limited scenario.
The acceleration time of electrons is (\citealt{2001RPPh...64..429M,2020pesr.book.....V})
\begin{equation}
    \tau_{acc}\approx 24\frac{\eta}{\delta}\sqrt{\frac{\varepsilon_0}{1~\rm{keV}}}\bigg( \frac{v_{sh}}{5000~\rm{km~s}^{-1}}\bigg)^{-2}\bigg( \frac{B}{100~\mu\rm{G} }\bigg)^{-\frac{3}{2}}~\rm{yr}\label{eq:tacc}
\end{equation}
where $\delta$ is a parameter which accounts for the energy dependence of the diffusion coefficient, typically ranging between 0.3 and 0.7 (see \citealt{2007ARNPS..57..285S}). Since we assume an energy independent (and rigidity independent) Bohm factor, we have $\delta=1$ by definition for our paper. The synchrotron cooling time is \citep{1994hea..book.....L}, 
\begin{equation}
    \label{eq:tsync}
    \tau_{sync} \approx 55 \bigg( \frac{\varepsilon_0}{1\mathrm{keV}}\bigg)^{-1/2}\bigg( \frac{B}{100\mathrm{\mu G}}\bigg)^{-3/2} \mathrm{yr}.
\end{equation}
We can derive the ratio $\tau_{acc}/\tau_{sync}$ which does not depend on the magnetic field, but depends on $v_{sh}$, $\varepsilon_0$, $\eta$ (the Bohm factor),  and $\delta$.
We estimated the lower limit of the ratio $\tau_{acc}/\tau_{sync}$ for each region by putting $\delta=\eta=1$ and propagating the errors associated with $v_{sh}$ and $\varepsilon_0$. 
Results are shown in Figure \ref{fig:ratio}.

The figure shows that, except for region 5, the acceleration time scale at the Bohm limit is always shorter than the synchrotron cooling time, i. e., in these regions loss-limited conditions ($\tau_{acc}/\tau_{sync}$ = 1)  can be achieved with $\eta\geq1$. 
In particular, in agreement with what we have discussed so far, loss-limited conditions are achieved with the values of $\eta_S$ and $\eta_N$ reported in Section \ref{sect:cut-vsh} (see also Figure \ref{fig:ecut}). In this case, we expect a steady synchrotron flux and no significant variations between 2006 and 2014. 

On the other hand, we find that in region 5 the ratio $\tau_{acc}/\tau_{sync}$ is significantly above the value of 1, whatever the Bohm factor.
This means that the acceleration time scale is way longer than the synchrotron cooling time, so the electrons in this regions are cooling faster than they are accelerating. 
We then expect the synchrotron flux of this region to decrease with time.
Figure \ref{fig:ratio} also shows a sort of trend for regions 4, 5 and 6, which exhibit higher ratios compared to other regions.
This could be suggestive that these regions may be entering a phase where the acceleration mechanisms are less efficient at counteracting particle losses.
Consequently, the synchrotron flux in these regions may decrease over future epochs and might be monitored by future observations.

\subsection{Flux Variability}
\begin{figure}
    \centering
    \includegraphics[trim={14pt 0 40pt 0},clip, width=\columnwidth]{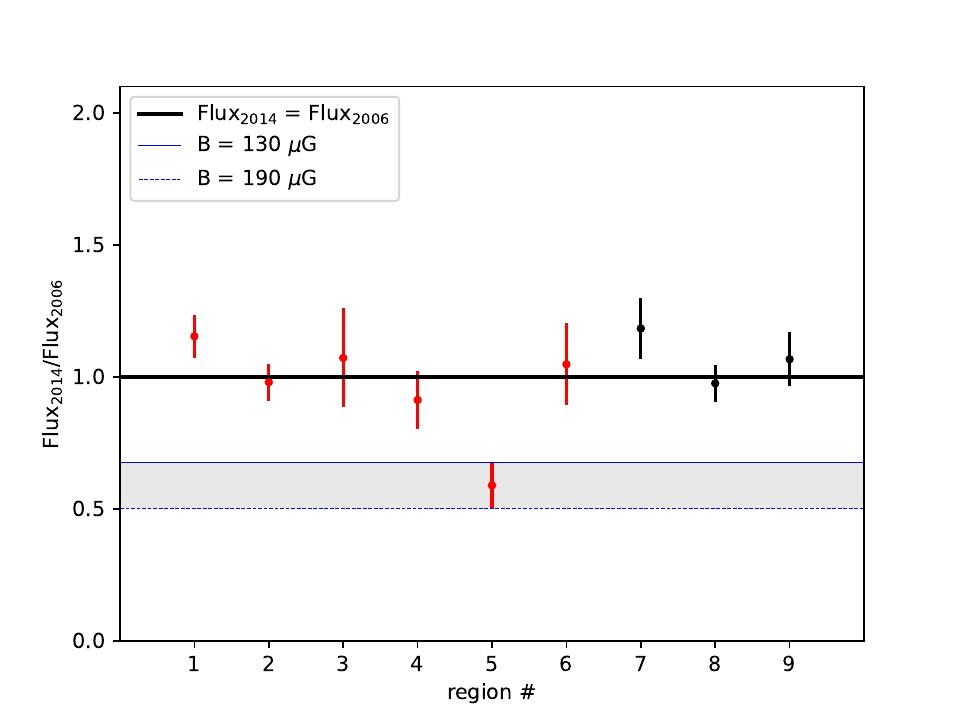}
    \caption{Ratio of the 2014 nonthermal flux ($4.1-6$ keV) to that measured in 2006 for northern regions (in red) and southern regions (in black).
    The blue lines represent the theoretical flux decrease expected for synchrotron losses with a magnetic field strength of $130\ \mathrm{\mu G}$ (solid line) and $190\ \mathrm{\mu G}$ (dashed line).
    }
    \label{fig:flux}
\end{figure}
In Figure \ref{fig:flux}, we illustrate the ratio of the synchrotron flux ($4.1-6$ keV) in 2014 to that in 2006 for the northern regions ($1-6$, in red) and in the southern regions ($7-9$, in black).  
Remarkably, except for region 5, the examined regions in the north do not show any significant decrease in flux. 
This confirms that these regions have remained within a loss-limited regime throughout the eight-year time base considered, as predicted in Section \ref{sect:times}.

Interestingly, the measurements of flux in region 5 show a significant decrease of the synchrotron radiation over this 8 year baseline.
This effect results from the synchrotron losses dominating over the acceleration efficiency, as predicted in Section \ref{sect:times} and in nice agreement with what we shown in Figure \ref{fig:ratio}.
As a further confirmation for this scenario, we show in Figure \ref{fig:flux} the comparison between the observed flux decrease with that expected from synchrotron \footnote{Theoretical decrease of the flux is calculated by numerical integration of the product of the `single-electron' emissivity and the momentum distribution of electrons taking into account their radiative losses in the magnetic field.} in a magnetic field of 130 $\mu$G and 190 $\mu$G (solid and dashed blue curves in Figure \ref{fig:flux}, respectively), showing the consistency between model and data points.

Conversely, region 1 and 7 show a barely significant ($\gsim 1\sigma$ level) increase in flux between the two epochs, which may suggest that in these regions $t_{acc}<t_{syn}$. Further investigations are necessary to confirm this latest point.


\subsection{Magnetic Field Strength from the Electron Cooling}
\label{sect:MF}

Figure \ref{fig:flux} shows that the flux decrease for region 5 agrees with theoretical predictions for electrons emitting in magnetic field with the strength $B\simeq 130-190\ \mathrm{\mu G}$ (blue lines). 
We here suggest another method to estimate $B$ in this region, by considering the decrease of the photon cutoff energy $\varepsilon_0$.

Electrons lose their energy through synchrotron radiation:  $\dot E=-AB^2E^2$ where $A=1/637$ in c.g.s. units.
By solving this differential equation for the final energy $E$, we obtain
\begin{equation}
    E(t)=\frac{E\rs{i}}{1+E\rs{i}/E\rs{f}(t)}
    \label{Kepler:Et}
\end{equation}
where $E\rs{i}$ is the initial energy of electrons (at the time $t\rs{i}$), $E\rs{f}=[AB^2(t-t\rs{i})]^{-1}$. 
The radiative losses are effective if the energy $E\rs{i}$ is about or larger than $E\rs{f}$. 
Electrons radiate most of synchrotron emission in photons with energy $\varepsilon_0=0.29c_2BE^2$ where $c_2=4.2\E{-8}$ in c.g.s. units. 
By relating $E\rs{i}$ and $E$ in Equation (\ref{Kepler:Et}) to the cut-off energy $\varepsilon_0$ at different times, we have an expression 
\begin{equation}
    B=1450\left(\frac{\varepsilon\rs{0,keV}^{-1/2}-\varepsilon\rs{0i,keV}^{-1/2}}{t\rs{yr}-t\rs{i,yr}}\right)^{2/3}\quad\mathrm{\mu G}
    \label{Kepler:epst}
\end{equation}
to estimate the strength of magnetic field in a region where the photon cut-off energy drops due to the radiative losses.
It yields $B=260\pm70\ \mathrm{\mu G}$ for region 5. 
This value agrees within errors with that derived from the flux decrease (refer to Figure~\ref{fig:flux}). 
The difference in estimation of the magnetic field strength could be due to a spatial orientation the magnetic field. 
Indeed, the synchrotron flux reflects the component of $B$ in the plane of the sky while the radiative losses of rapidly isotropized electrons are sensitive to the total $B$.

We note that Equation \ref{Kepler:epst} assumes that no electrons with energy $\geq E$ are supplied by the diffusive shock acceleration. 
If this is not the case, the estimated value of $B$ should be considered as the lower bound for the magnetic field strength in the region. 

\subsection{A comparison with radio polarization}
\label{subsect:polariz}
The measurement of polarization plays a significant role in SNRs as it serves as an indicator of the ordering of the magnetic field.
Taking advantage of the radio fractional polarization map we can test the scenario we have proposed for Kepler's SNR.

\begin{figure}
    \centering
    \includegraphics[width=\columnwidth]{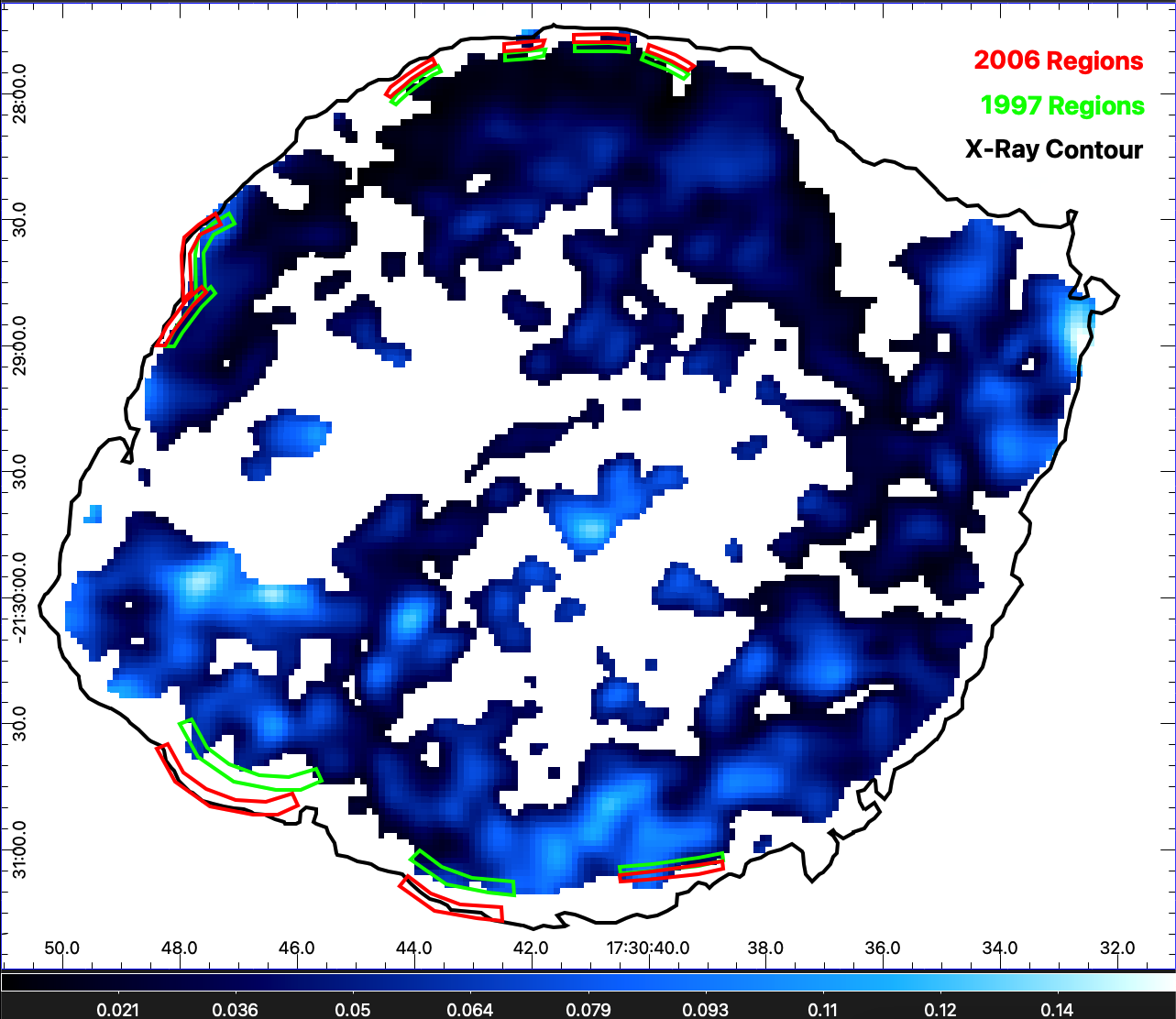}
    \caption{Polarization fraction at 1.4 GHz from \cite{2002ApJ...580..914D}. 
    The colorbar is in linear scale.
    The black contours overlaid indicate the \textit{Chandra}/ACIS image for the year 2000. 
    The red polygons are the same as Figure \ref{fig:reg}.
    The green polygons are the extraction regions (shifted to take in account the proper motion) to measure the spectral index and polarization fraction. 
    }
    \label{fig:perpol}
\end{figure}
Figure \ref{fig:perpol} presents the 1.4 GHz fractional polarization map from \citet{2002ApJ...580..914D}.
In the map, the north shows a polarization fraction which is, on average, lower than in the south.
Since a lower value of polarization fraction is associated with a more turbulent magnetic field, this results is an 
additional proof that points toward a scenario in which the electron acceleration in the north is enhanced by a turbulent magnetic field, generated in the interaction of the shock with the dense CSM.

A similar pattern was found by \citet{2022ApJ...938...59S} in RCW 86 (in the southwestern region where the remnant is interacting with a dense cloud).
In particular, they retrieved a low Bohm factor in the region where radio observations show a low degree of polarization \citep{2001ApJ...546..447D}.
\citet{2021Ap&SS.366...58S} propose that the observed effect occurs not only due to the density of the CSM but also its clumpiness.
\cite{2023PASJ...75.1344B} comparing the southwest rim in RCW 86 of \cite{2022ApJ...938...59S} with the northwest rim which show opposite behaviors, also suggest that this difference may be due to the clumpiness of dense material interacting with the shock.

We calculated the polarization fraction and the radio index from the data of \cite{2002ApJ...580..914D} in the nine regions shown in Figure~\ref{fig:reg}. The regions corresponds to regions $1-9$ of Figure \ref{fig:reg}, but are shifted inwards to account for the expansion of the SNR in the period between the X-ray and radio observations (which date back to 1997). 
The measurements are shown in Table~\ref{tab:alphaPi}: the polarization fraction in the North (regions 3-6) is about 3 times smaller than in the South (regions 7-9). 

One may use these data for an independent estimate of the difference between the Bohm factor in the North and in the South, for the radio emitting electrons. 
\citet{2016MNRAS.459..178B} generalized the classic synchrotron theory to cases where electrons emit in the magnetic field with ordered and disordered components. 
Equation (31) in this reference relates the radio index and the ratio ${\delta B_\perp}/{B_\perp}$ to the polarization fraction (the index $\perp$ refers to the components in the plane of the sky). 
This relation yields  ${\delta B_\perp}/{B_\perp}=4.20\pm0.01$ and $2.30\pm 0.14$ for the North and South respectively for the radio data at 1.4 GHz and $2.9\pm0.01$ and $2.00\pm 0.07$ for the data at 4.8 GHz.
Next, the relation $\eta\approx\left({\delta B}/{B}\right)^{-2}$ may be used to estimate $\eta$ from ${\delta B}/{B}$. In this expression, the strengths of the turbulent $\delta B$ and of the ordered $B$ components refer to the three-dimensional vectors which are related to the projected ones as $\delta B_\perp=\delta B/\sqrt{3}$, $B_\perp=B\sin{\varphi}$ with $\varphi$ being the angle between the line of sight and the vector $\mathbf{B}$. 
We do not know the actual three-dimensional orientation of $\mathbf{B}$ in our regions of interest. 
Therefore, by assuming $\varphi$ to be the same in all the regions, we derive a ratio between the Bohm parameter of southern regions ($\eta_S$) and the Bohm parameter of northern regions ($\eta_N$) as $\eta_S/\eta_N= 3.3\pm0.4$ and $2.2\pm0.5$ for the radio data at 1.4 and 4.8 GHz respectively.
These values correspond within errors to that obtained from the X-ray data analysis.

The data on polarization fraction reveal that the disordered component $\delta B$ of magnetic field is higher than the ordered component $B$ in our regions. Figure~11 of \citet{2002ApJ...580..914D} shows that the magnetic field vectors are quite ordered over the whole SNR surface, being mostly aligned with a radial orientation. Such situation may have place if the high degree of disorder is on scales that are not resolved by the angular resolution of the radio data we used. Another possibility could be if the turbulence is isotropic and the randomly oriented $\delta B$ vectors cancel out each other within the regions and relevant lines of sight leaving the only the ordered component $B$ visible in the polarization vector maps.

\begin{table}
\centering
    \caption{The radio index $\alpha$ between 1.4 and 4.8 GHz and the polarization fraction $\Pi$ at 1.4 and 4.8 GHz from the radio data of \cite{2002ApJ...580..914D}. Errors shown are 1$\sigma$ statistical only. %
    }
\begin{tabular}{cccc}
\hline\hline         
    region&	$\alpha$&	$\Pi_{1.4}$, \% & $\Pi_{4.8}$, \% \\
\hline
1&	$0.680\pm0.009$&	$4.1\pm0.5$ &   $7.3\pm0.5$\\
2&	$0.670\pm0.007$&	$7.4\pm1.2$ &   $13.6\pm2.7$\\
3&	$0.690\pm0.006$&	$2.1\pm0.2$ &   $4.5\pm0.8$\\
4&	$0.680\pm0.002$&	$1.5\pm0.1$ &   $4.6\pm0.2$\\
5&	$0.670\pm0.003$&	$2.7\pm0.2$ &   $3.0\pm0.7$\\
6&	$0.680\pm0.002$&	$1.4\pm0.2$ &   $3.5\pm0.8$\\
7&	$0.660\pm0.019$&	$6.4\pm2.5$ &   $9.0\pm3.2$\\
8&	$0.660\pm0.009$&	$6.2\pm1.5$ &   $5.6\pm1.4$\\
9&	$0.690\pm0.022$&	$5.3\pm1.5$ &   $9.3\pm2.9$\\
\hline
\end{tabular}
\label{tab:alphaPi}
\end{table}

\section{Conclusions}\label{sec:con}
We analyzed different \textit{Chandra}/ACIS archive observations of the Kepler's SNR in two separate epochs: 2006 and 2014.
Our analysis has added some significant findings to our previous study \citep{2022ApJ...935..152S}, regarding particle acceleration and synchrotron emission in Kepler's SNR.
Firstly, our research has confirmed the existence of two distinct regimes of particle acceleration, with electron acceleration proceeding closer to the Bohm limit in the northern part of the remnant than in the freely expanding southern part. 
This strongly supports the scenario in which the interaction between the shock and dense CSM in the northern regions results in the amplification of magnetic field turbulence that speed-up the acceleration process.
The analysis of the polarization fraction in Kepler's SNR provides a further compelling evidence that the acceleration process is influenced by the magnetic field's turbulent nature. 
The lower polarization fraction observed in the northern region, in fact, suggests the presence of an enhancement of the magnetic field turbulence therein.

While the reduction of the Bohm factor in the North speeds up the acceleration process, the interaction with the dense CSM significantly decelerates the shock, thus increasing the acceleration time. 
We find that the two effects compensate to each other in the majority of the regions analyzed, and the electron acceleration proceeds steadily in the loss-limited regime. 
However, in one region (characterized by a very low shock speed) we find that the radiative losses dominate over the acceleration process and we report a gradual decrease in the synchrotron emission together with a decrease in the cutoff photon energy.

This provides the evidence of a fading synchrotron emission in the northern part of Kepler's SNR, which is in agreement with earlier results by \citep{1988ApJ...330..254D}. 
The drop of the flux and of the cutoff energy allows us to estimate the local strength of the magnetic field ($B\sim130-190~\mu$G).

Overall, our study provides a coherent and comprehensive understanding of the electron acceleration in Kepler's SNR. 
The role played by the CSM interacting with the shock in affecting the particle acceleration and synchrotron emission has been elucidated, contributing to a deeper comprehension of this fascinating celestial object. 

\section*{Acknowledgements}
V.S., S.O. and M.M. acknowledge financial contribution from the PRIN MUR “Life, death and after-death of massive stars: reconstructing the path from the pre-supernova evolution to the supernova remnant” funded by European Union - Next Generation EU.
M.M. and V.S. acknowledge financial support by the INAF mini-grant “X- raying shock modification in supernova remnants"
FB, MM, OP, and SO acknowledge financial contribution from the INAF Theory Grant ``Supernova remnants as probes for the structure and mass-loss history of the progenitor systems''.
O.P. acknowledges the OAPa grant number D.D.75/2022 funded by Direzione Scientifica of Istituto Nazionale di Astrofisica, Italy. This project has received funding through the MSCA4Ukraine project, which is funded by the European Union. Views and opinions expressed are however those of the authors only and do not necessarily reflect those of the European Union. Neither the European Union nor the MSCA4Ukraine Consortium as a whole nor any individual member institutions of the MSCA4Ukraine Consortium can be held responsible for them.
This paper employs a list of Chandra datasets, obtained by the Chandra X-ray Observatory, contained in~\dataset[DOI: 10.25574/cdc.267]{https://doi.org/10.25574/cdc.267}.
\appendix
\section{Spectra}
\begin{figure*}[h!]
    \centering
    \includegraphics[width=0.3\textwidth]{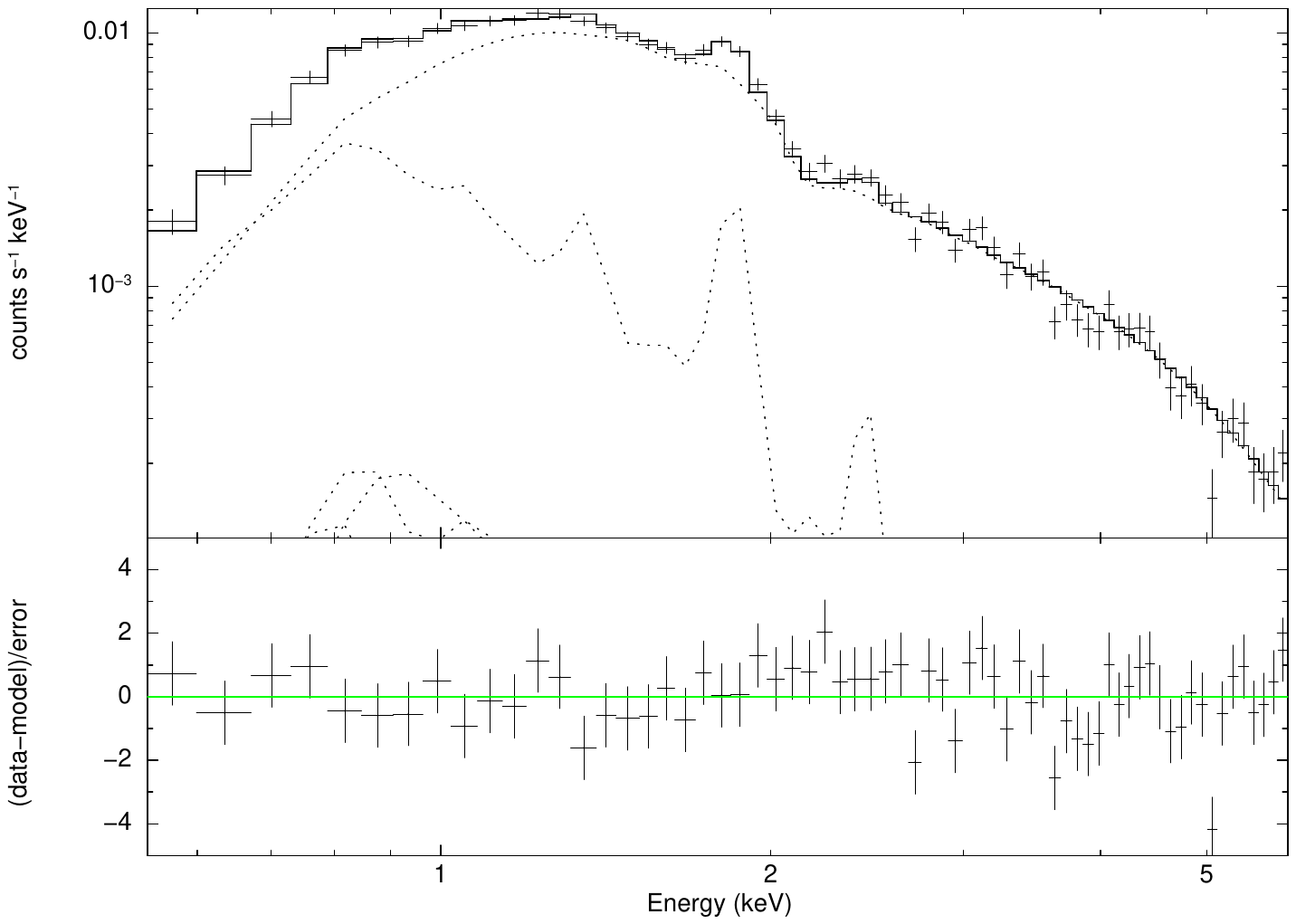}
    \includegraphics[width=0.3\textwidth]{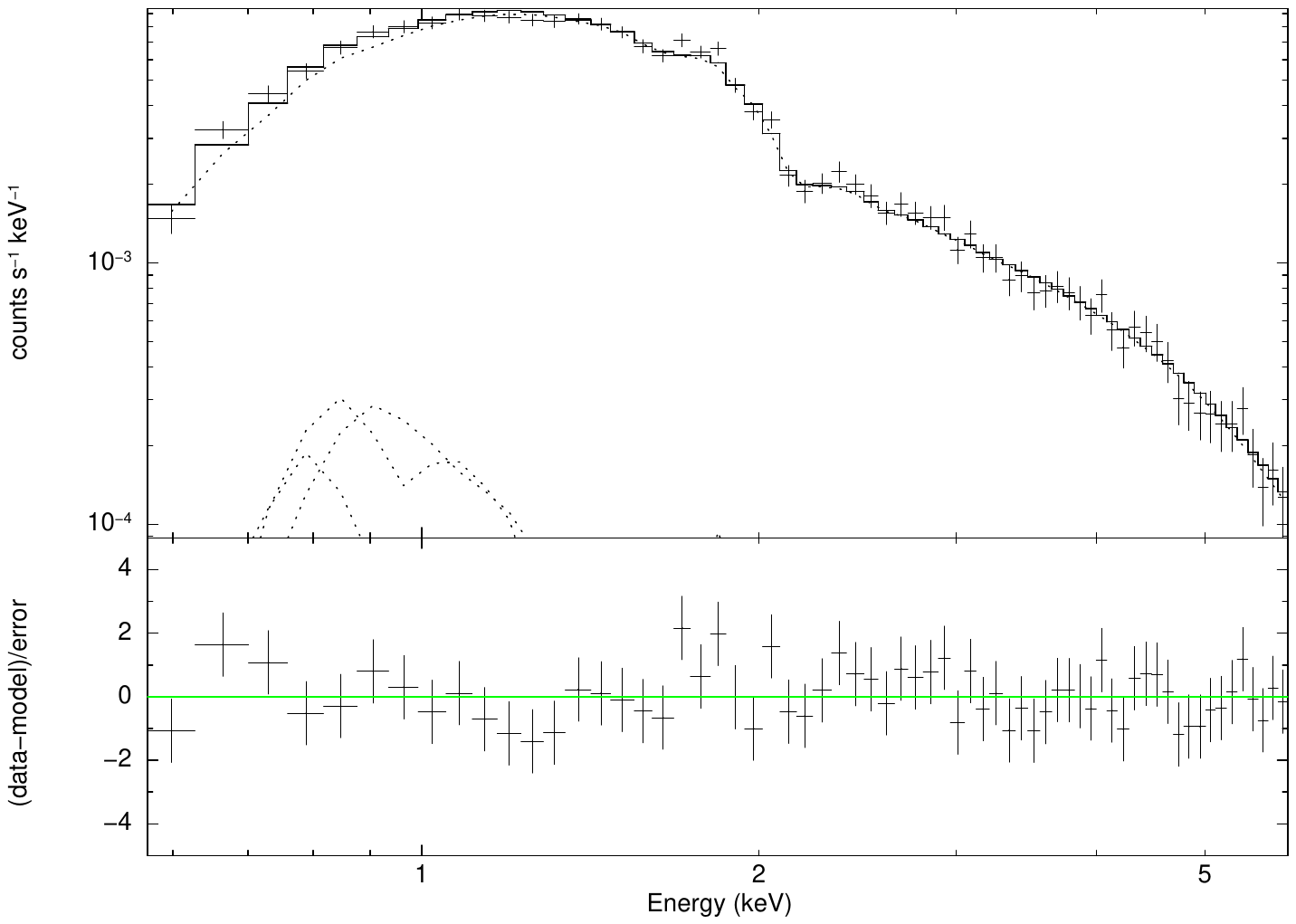}
    \includegraphics[width=0.3\textwidth]{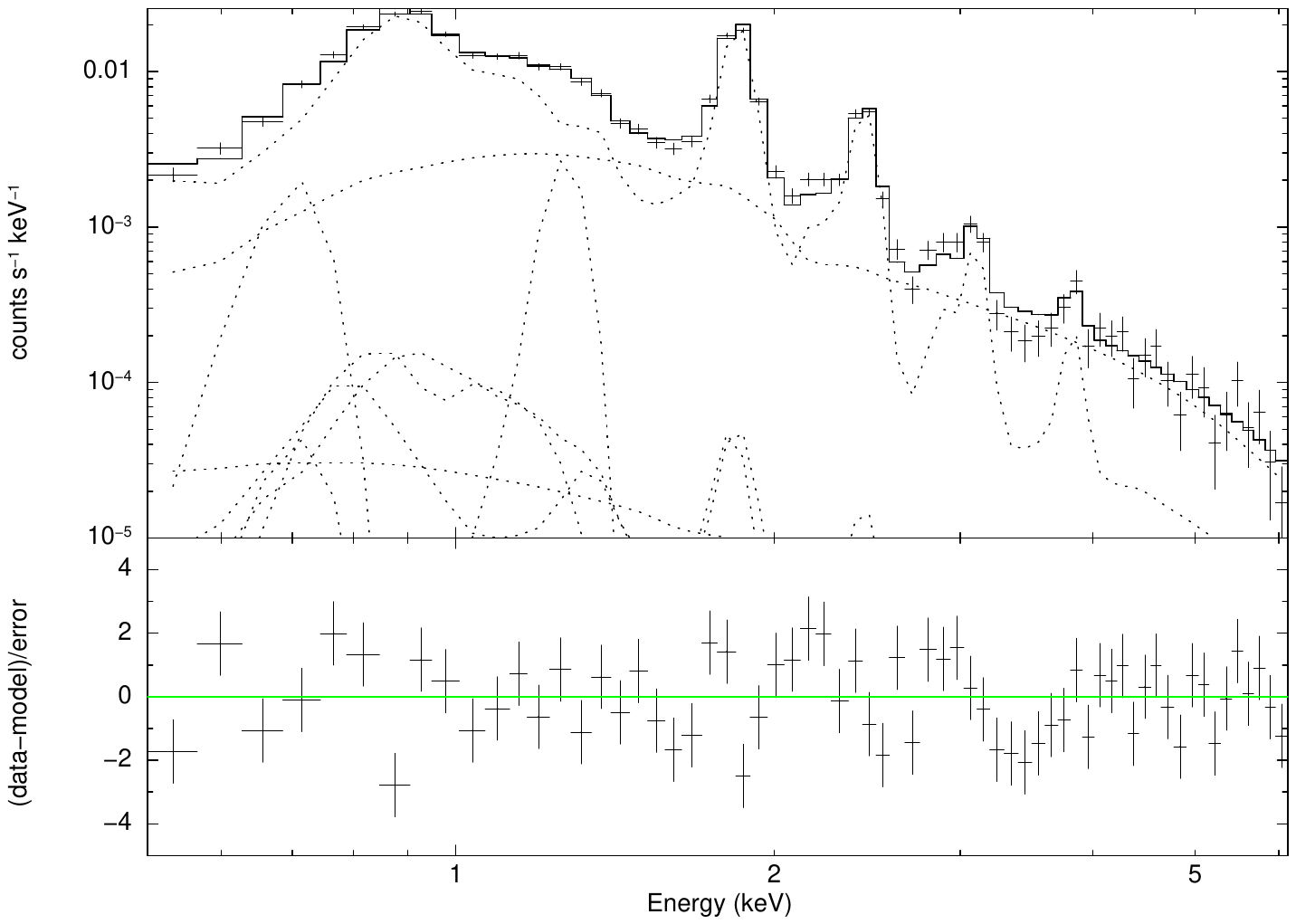}
    \includegraphics[width=0.3\textwidth]{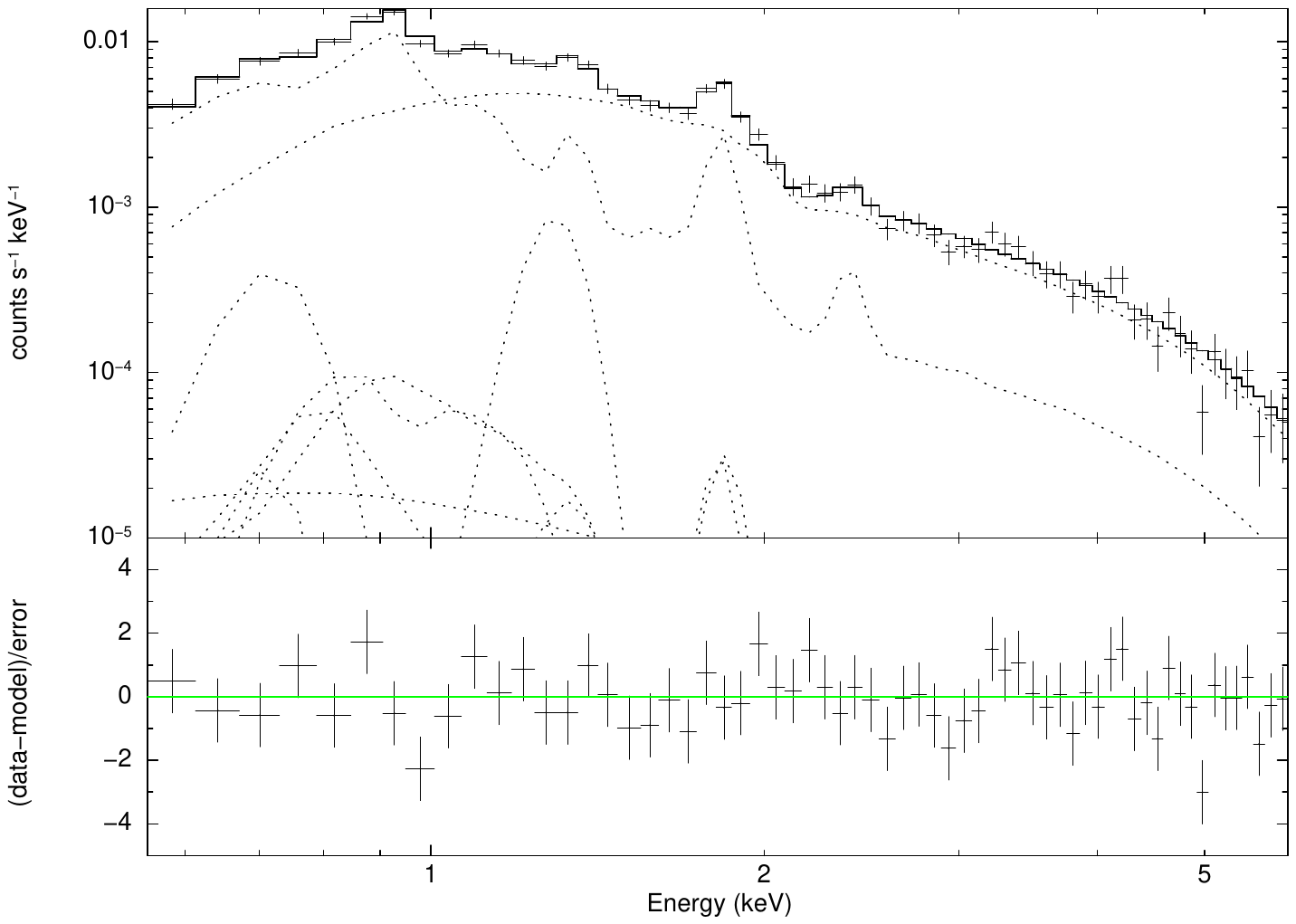}
    \includegraphics[width=0.3\textwidth]{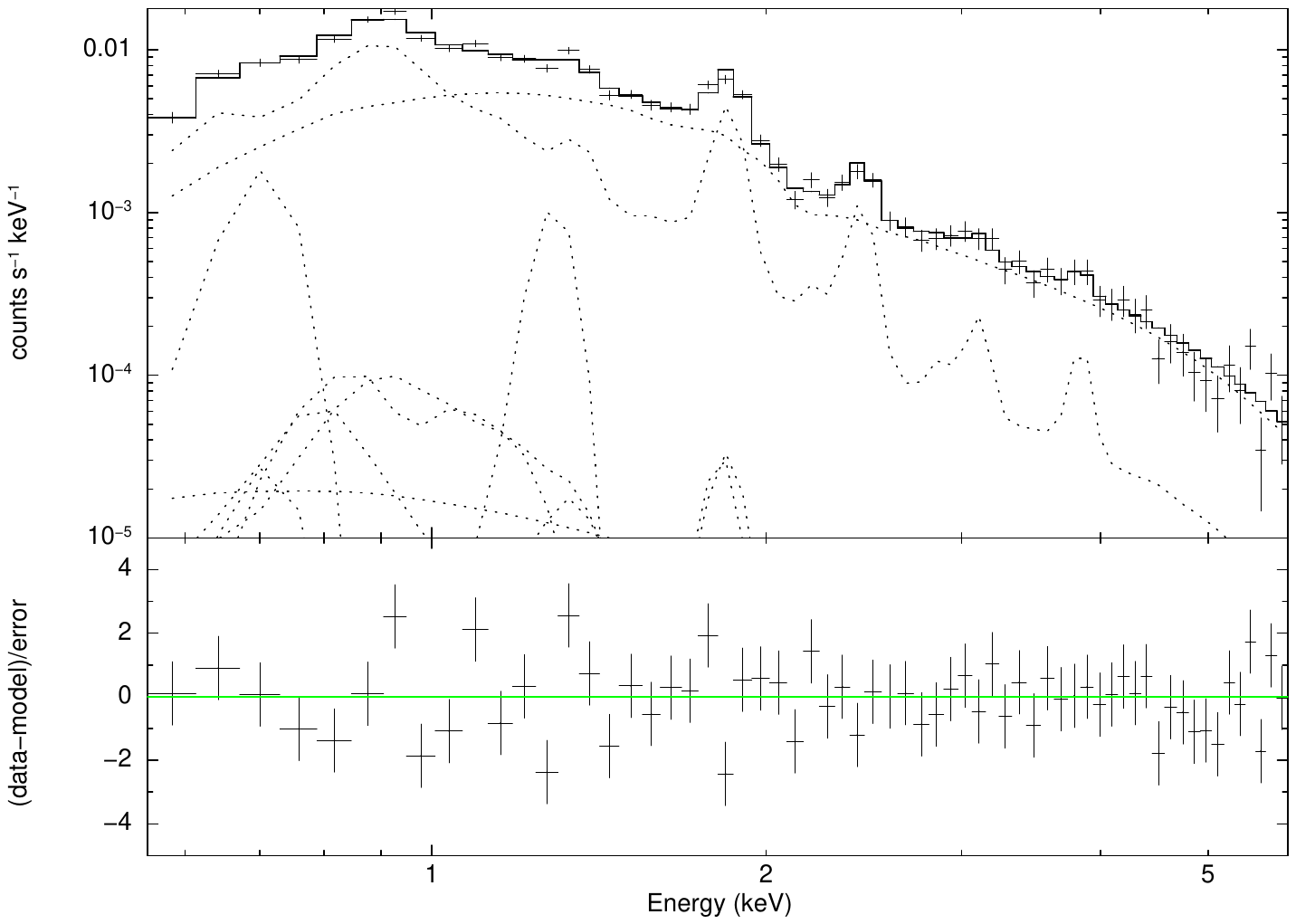}
    \includegraphics[width=0.3\textwidth]{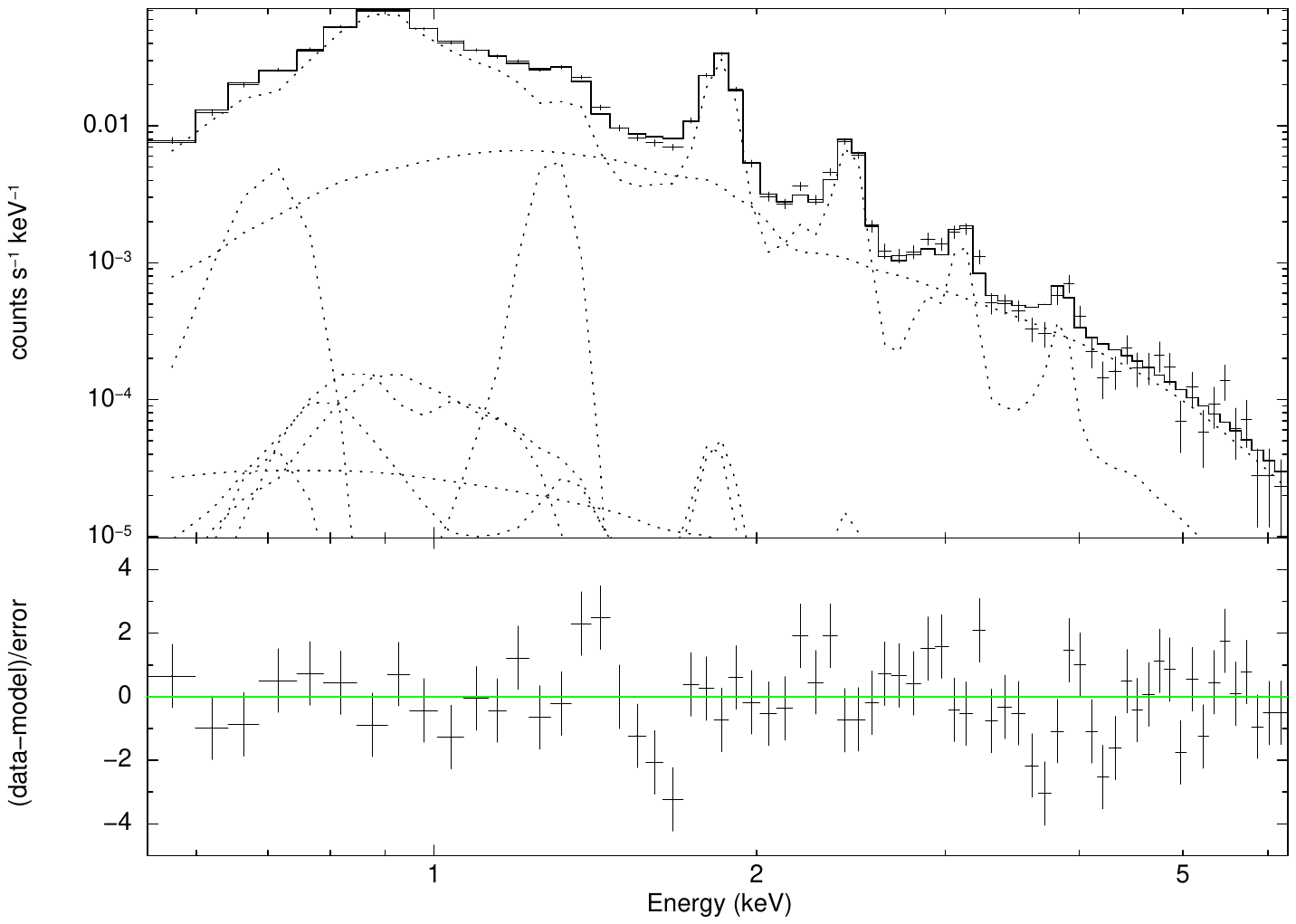}
    \includegraphics[width=0.3\textwidth]{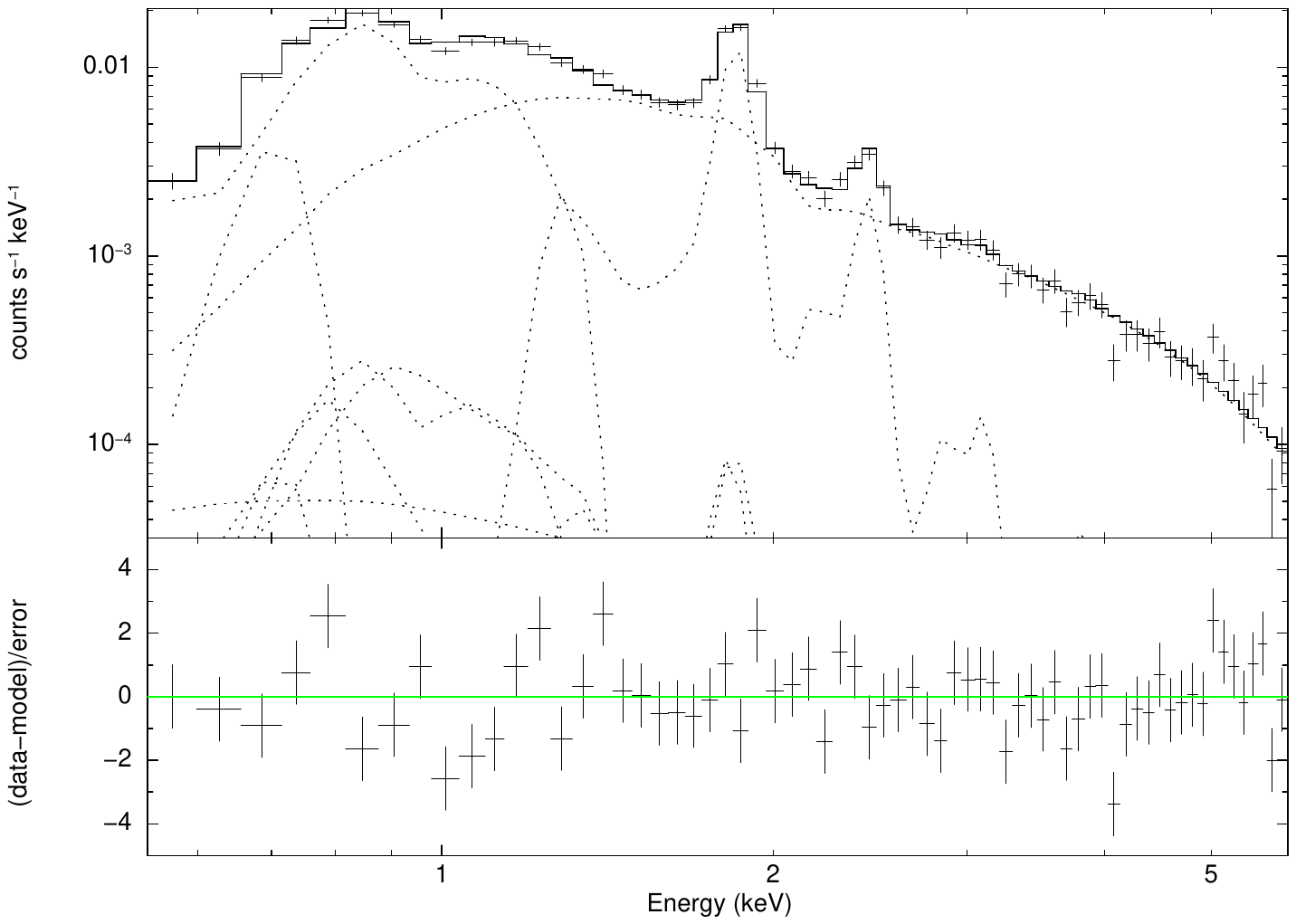}
    \includegraphics[width=0.3\textwidth]{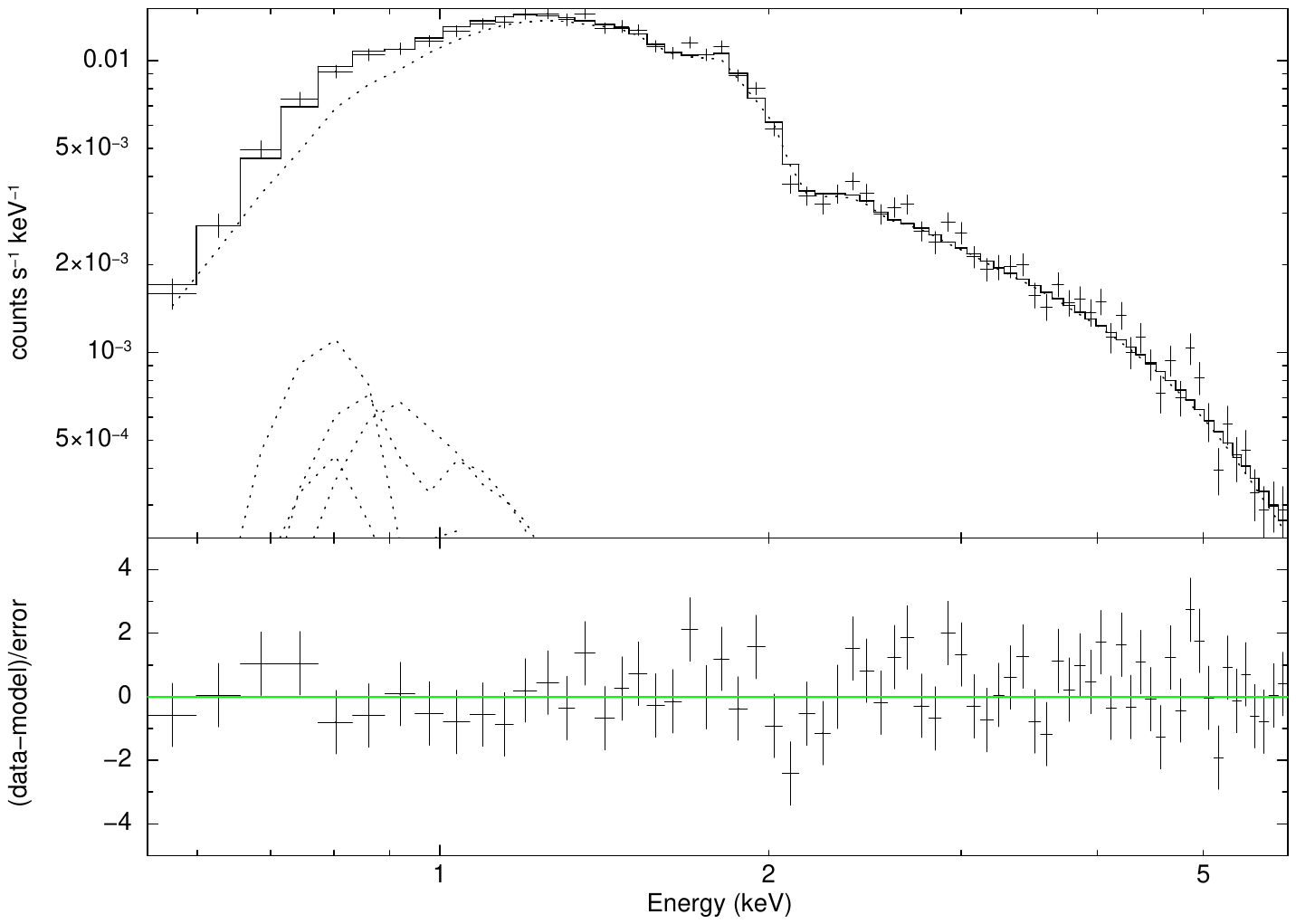}
    \includegraphics[width=0.3\textwidth]{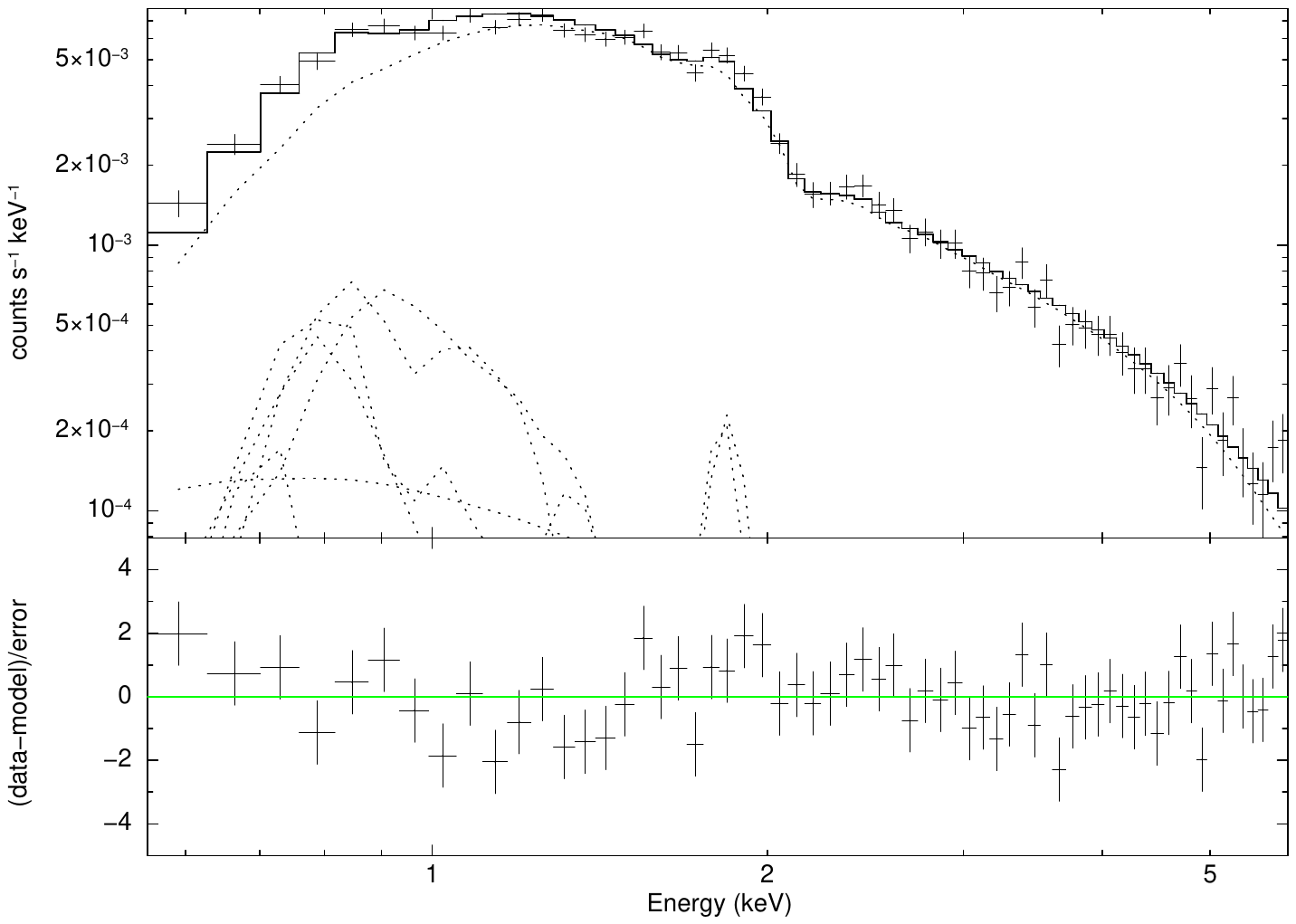}
    \caption{Spectra extracted from the regions in Figure \ref{fig:reg} with best-fit models and residuals for the year 2006}
    \label{fig:2006spectra}
\end{figure*}
\begin{figure*}[h!]
    \centering
    \includegraphics[width=0.3\textwidth]{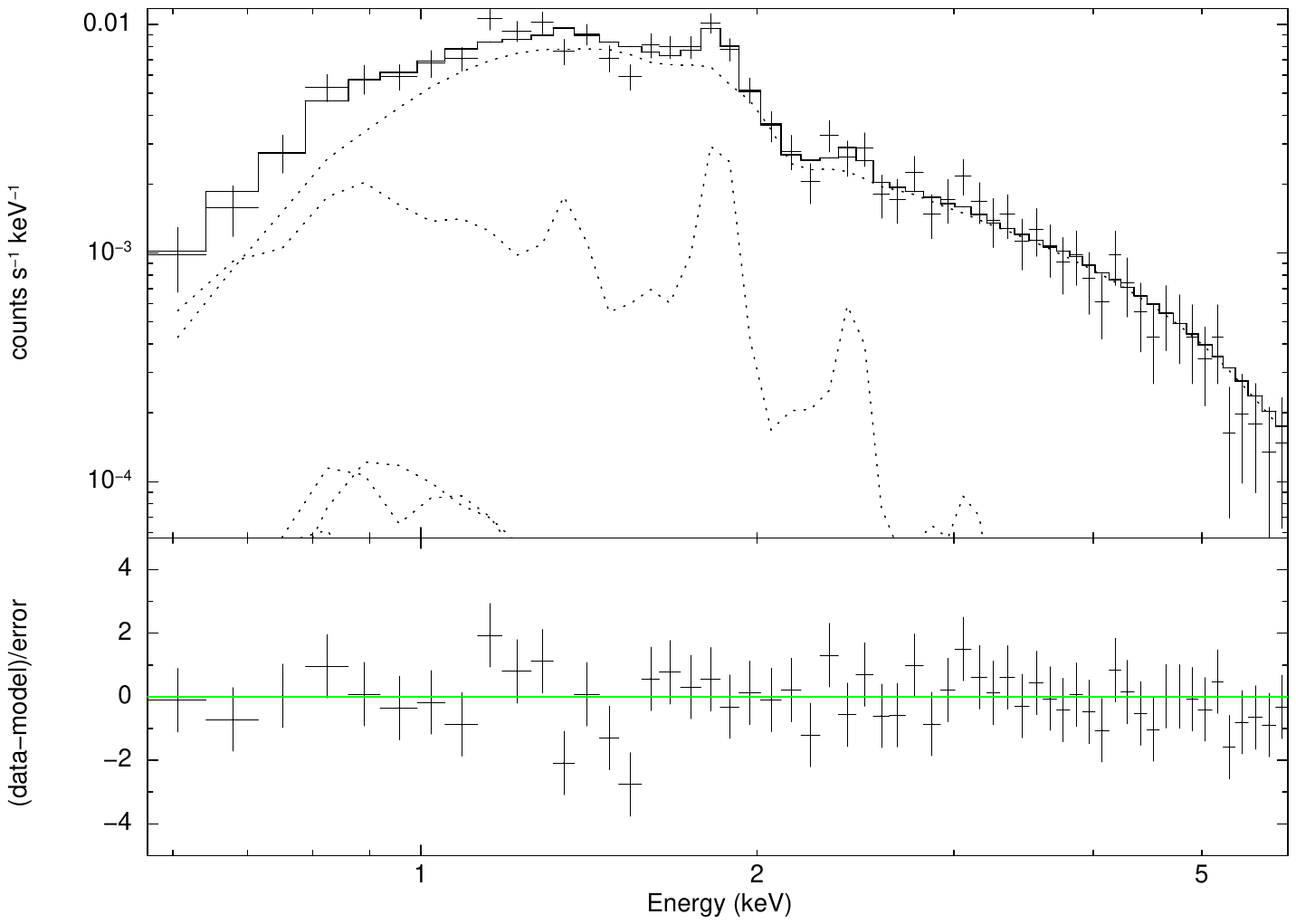}
    \includegraphics[width=0.3\textwidth]{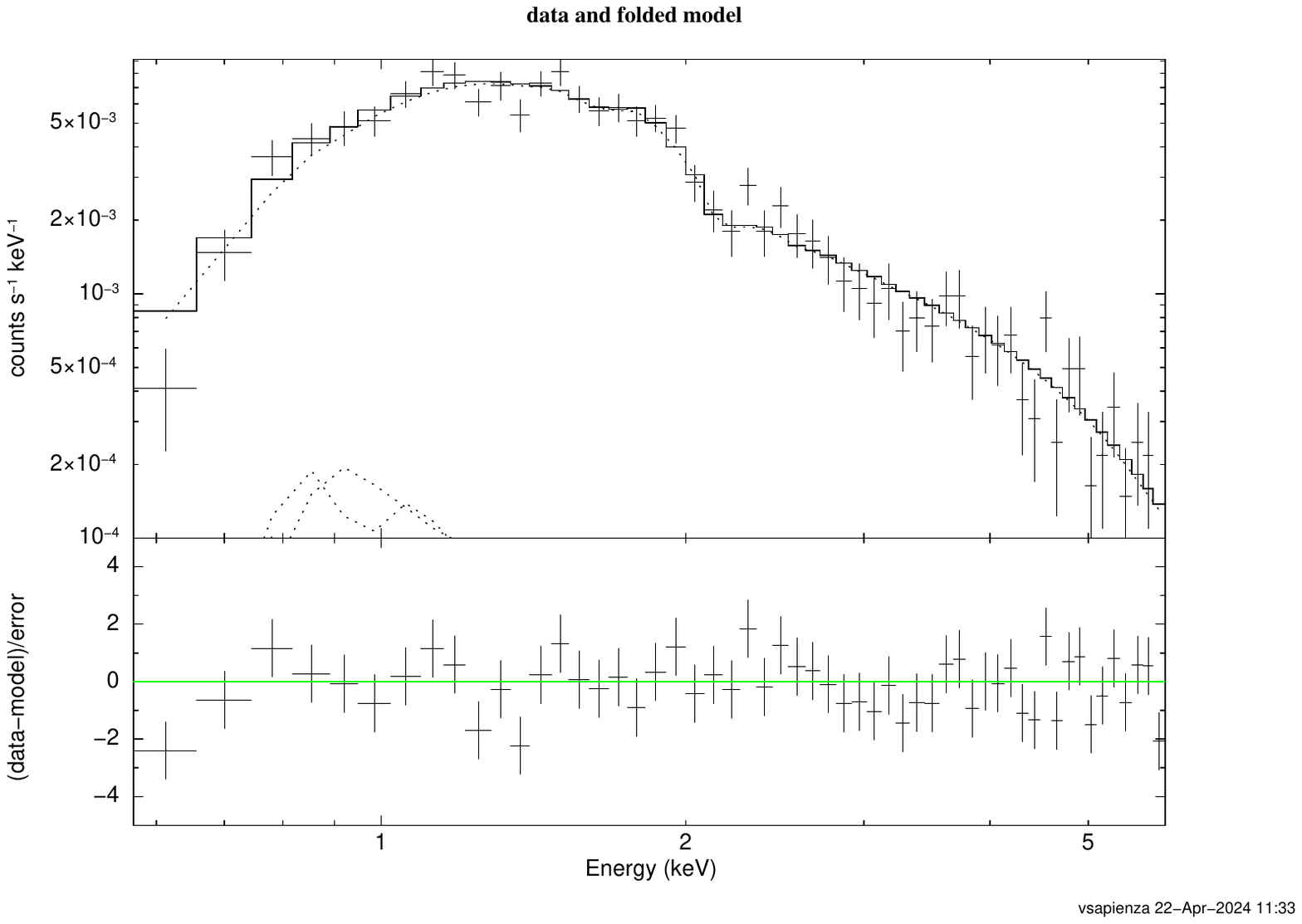}
    \includegraphics[width=0.3\textwidth]{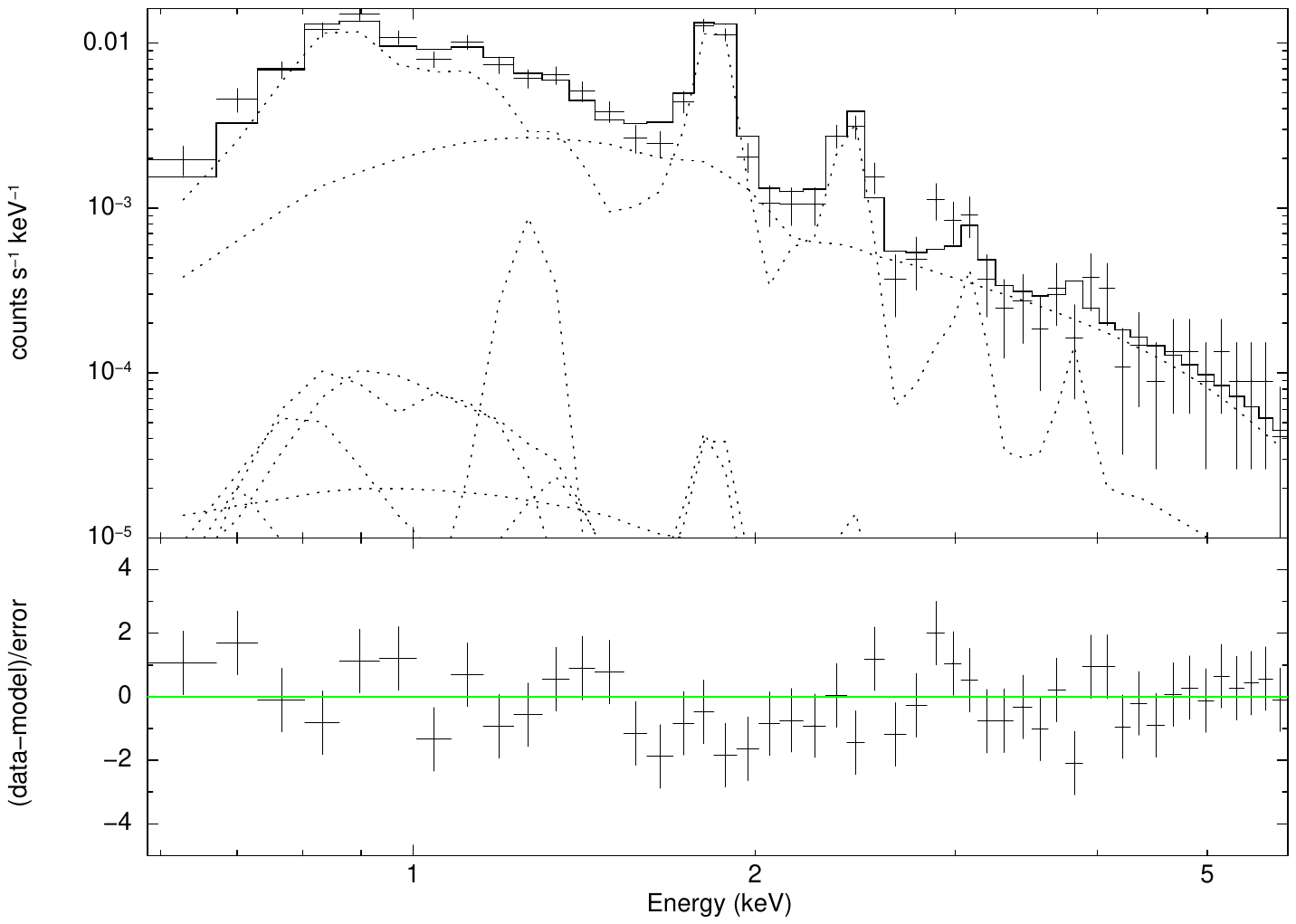}
    \includegraphics[width=0.3\textwidth]{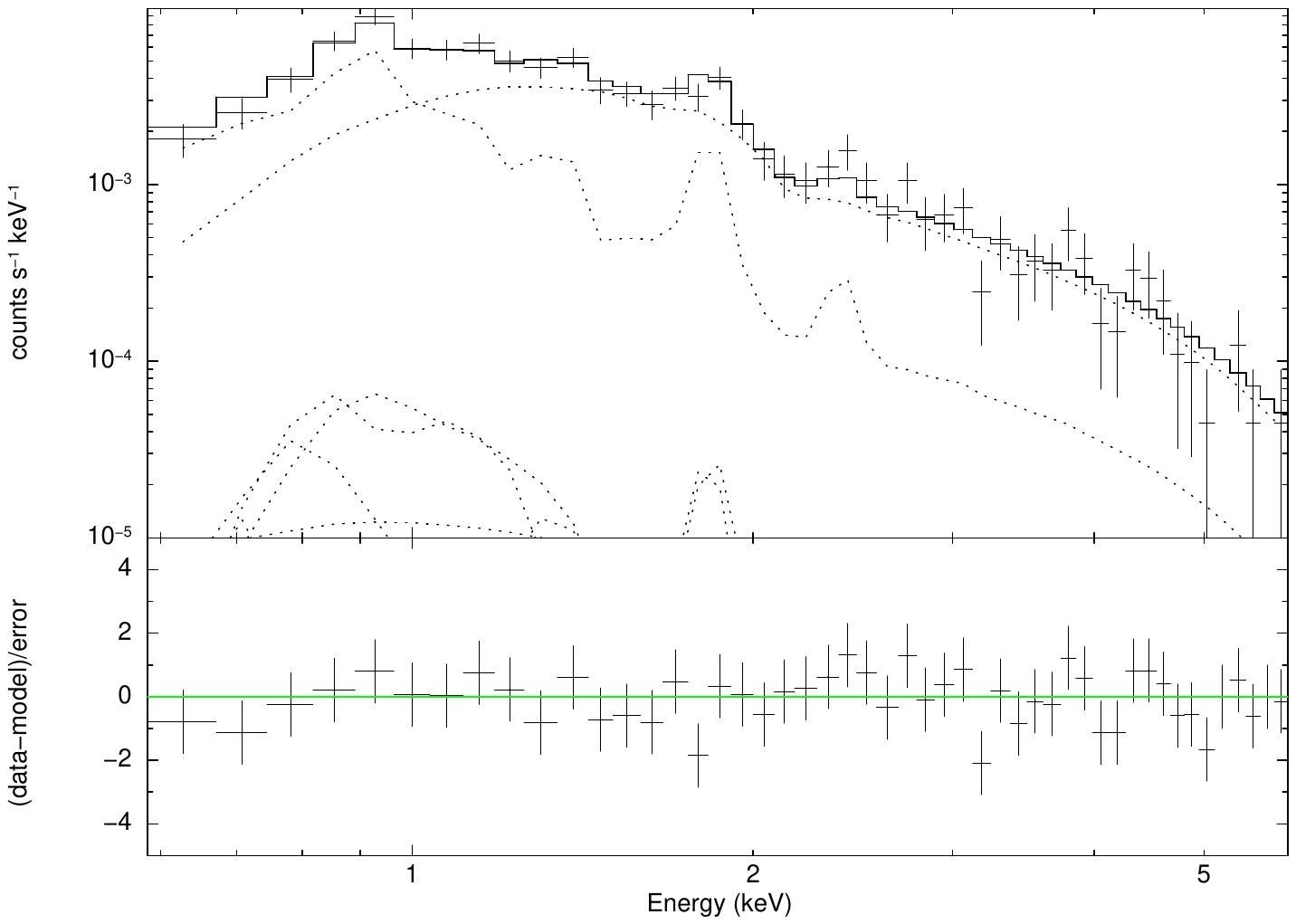}
    \includegraphics[width=0.3\textwidth]{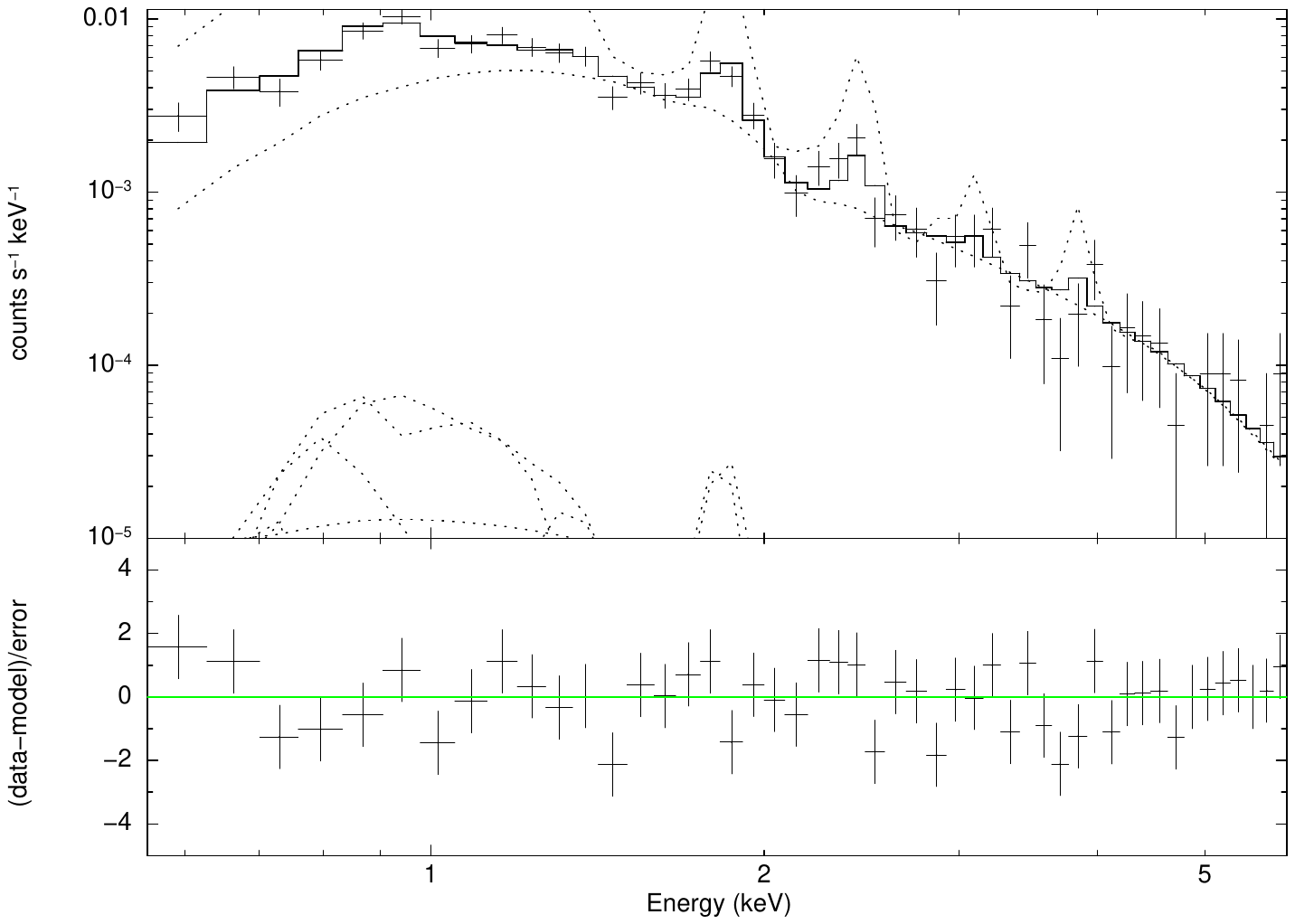}
    \includegraphics[width=0.3\textwidth]{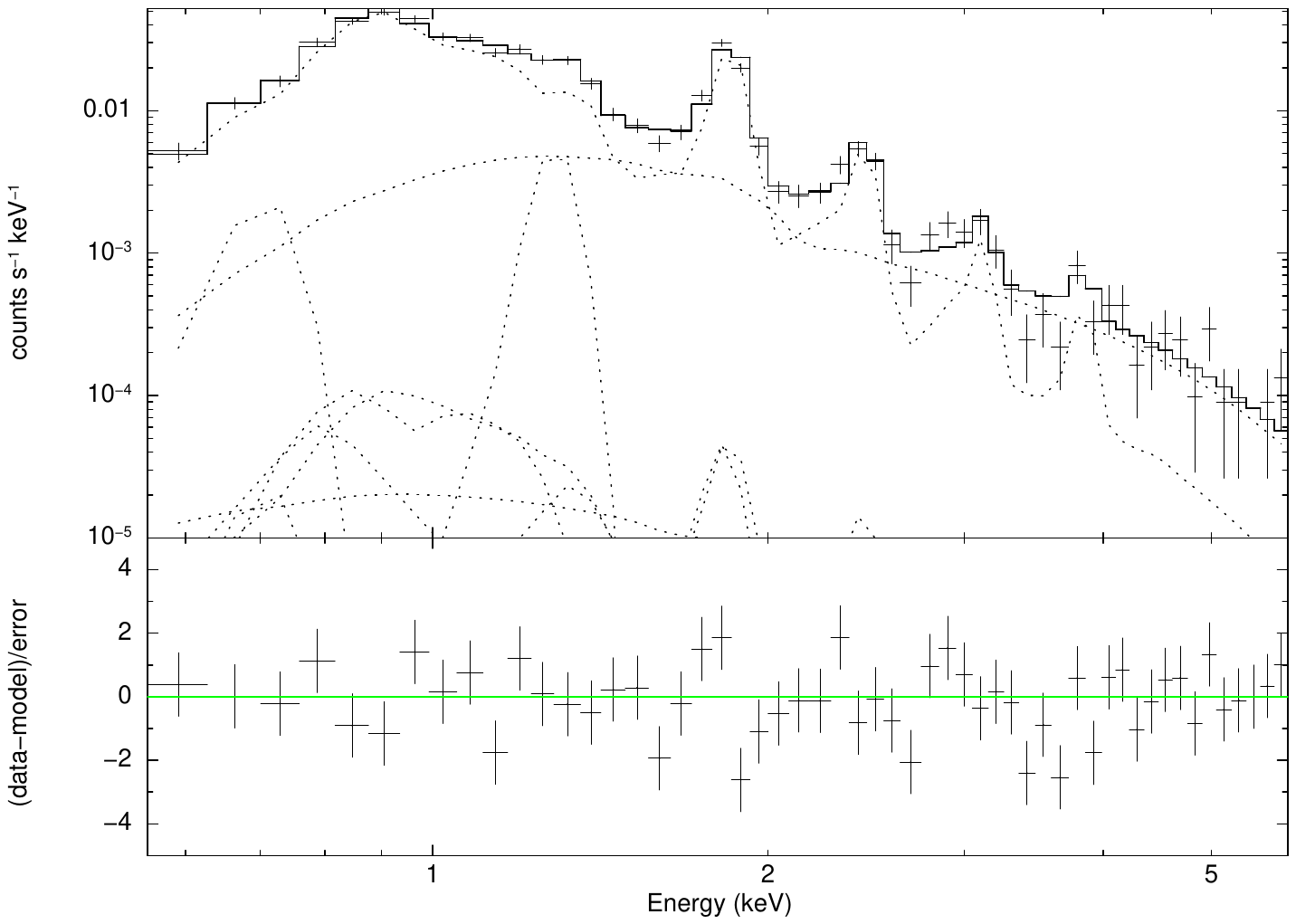}
    \includegraphics[width=0.3\textwidth]{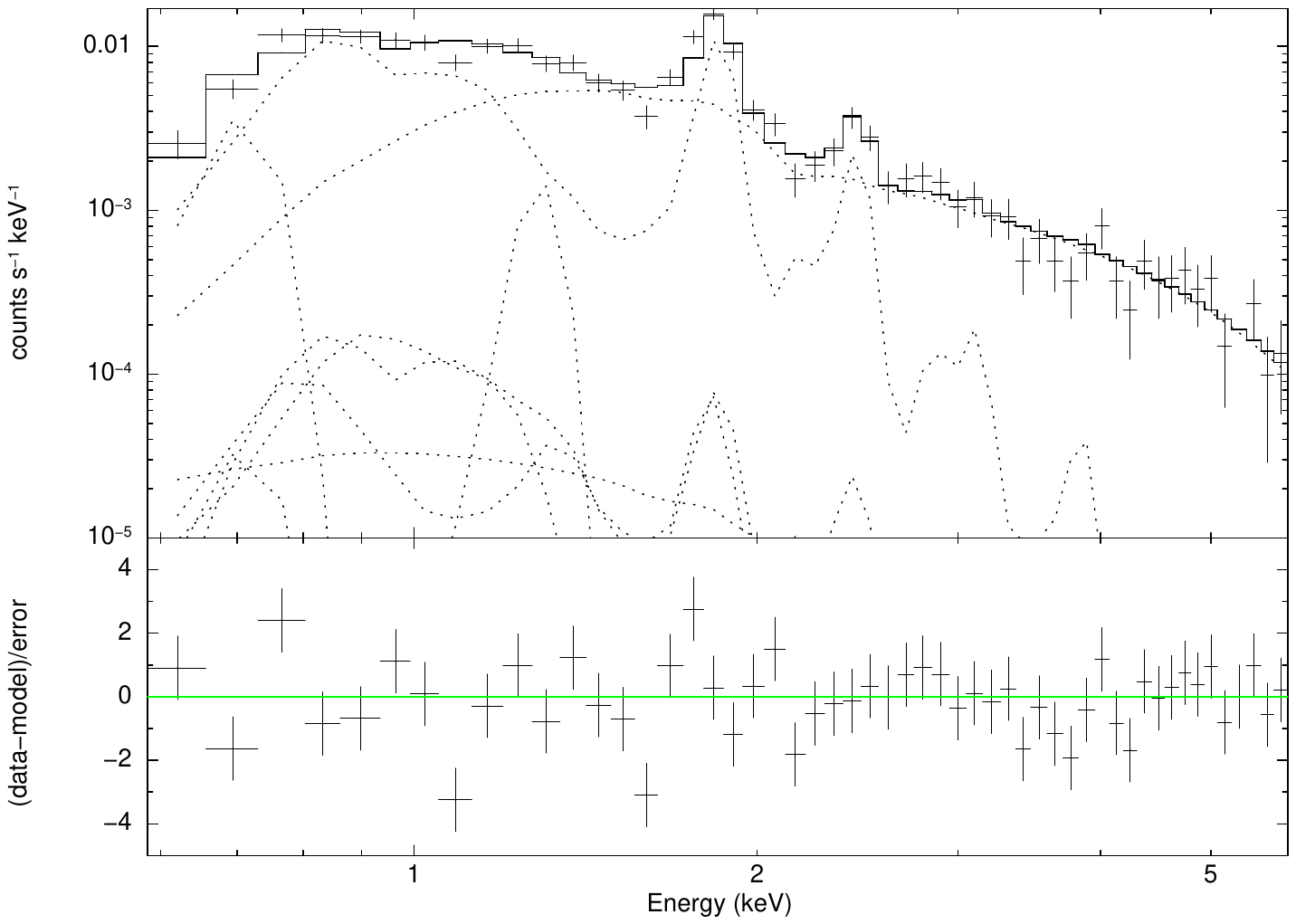}
    \includegraphics[width=0.3\textwidth]{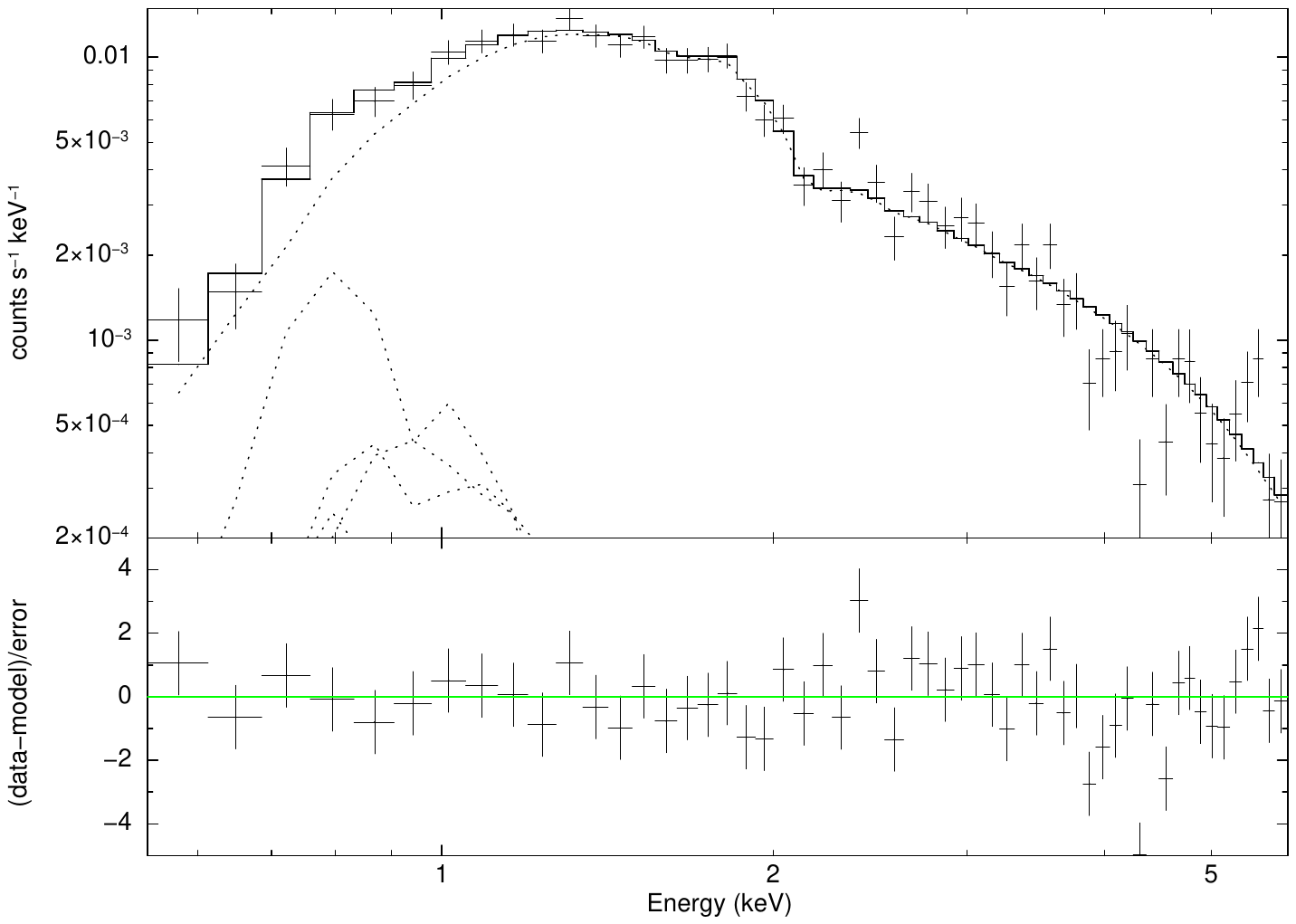}
    \includegraphics[width=0.3\textwidth]{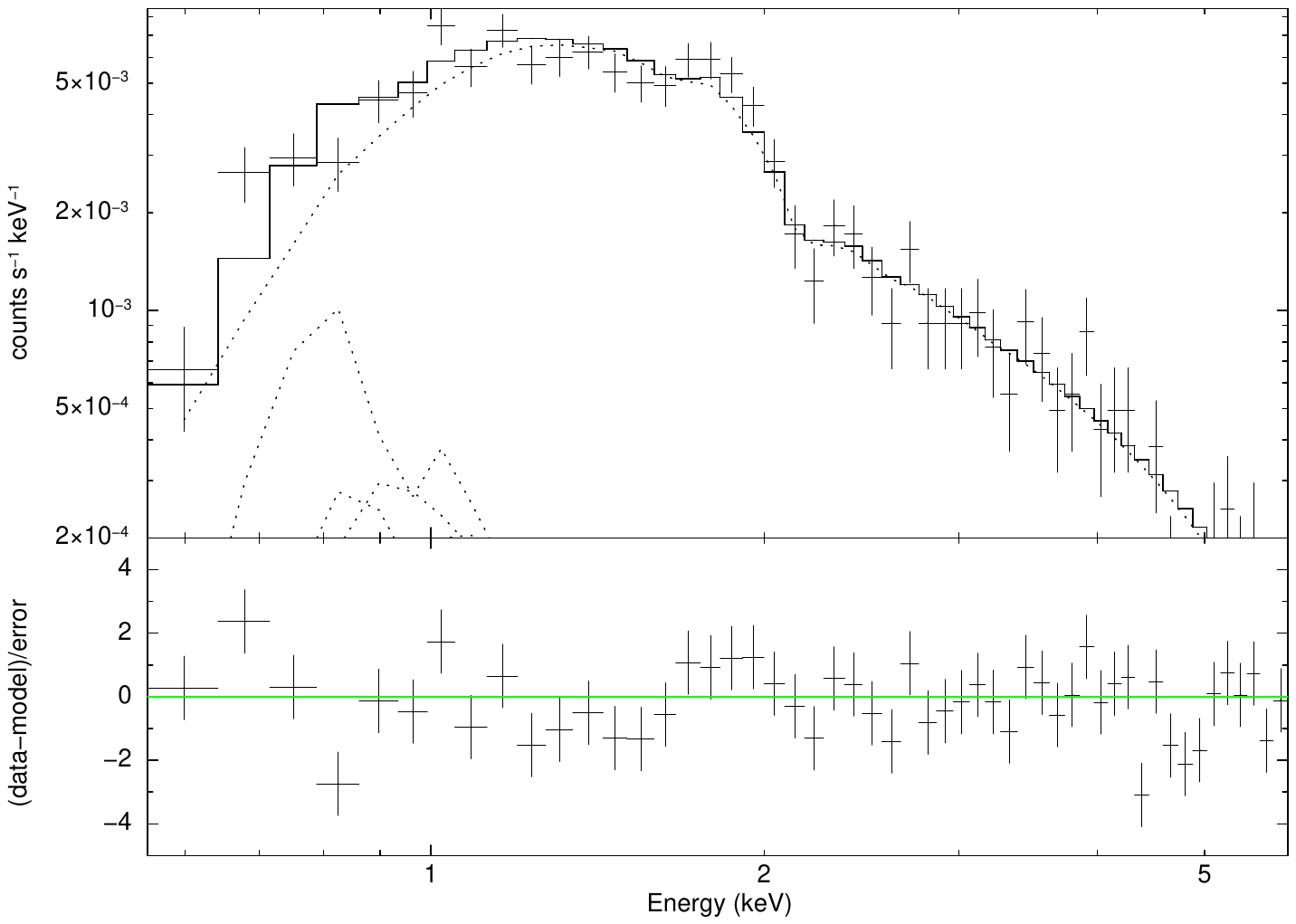}
    \caption{Spectra extracted from the regions in Figure \ref{fig:reg} with best-fit models and residuals for the year 2014}
    \label{fig:2014spectra}
\end{figure*}
\newpage
\bibliography{sample631}{}
\bibliographystyle{aasjournal}



\end{document}